\documentclass[aps,pre,twocolumn,groupedaddress,showpacs,floatfix]{revtex4}
\usepackage{amsmath}
\usepackage{amssymb}
\usepackage{graphicx}
\usepackage[bf]{subfigure}

\begin{document}

\title{Analyzing Trails in Complex Networks}
\author{Luciano da Fontoura Costa}\email{luciano@if.sc.usp.br}
\author{Francisco A. Rodrigues}\email{francisco@ifc.usp.br}
\author{Gonzalo Travieso}\email{gonzalo@ifsc.usp.br}

\affiliation{Instituto de F\'{\i}sica de S\~ao Carlos, Universidade de
  S\~ ao Paulo, S\~{a}o Carlos, SP, PO Box 369, 13560-970, phone +55
  16 3373 9858,FAX +55 16 3371 3616, Brazil, }

\begin{abstract}

  Even more interesting than the intricate organization of complex
  networks are the dynamical behavior of systems which such structures
  underly.  Among the many types of dynamics, one particularly
  interesting category involves the evolution of trails left by moving
  agents progressing through random walks and dilating processes in a
  complex network. The emergence of trails is present in many
  dynamical process, such as pedestrian traffic, information flow and
  metabolic pathways. Important problems related with trails include
  the reconstruction of the trail and the identification of its
  source, when complete knowledge of the trail is missing. In addition,
  the following of trails in multi-agent systems represent a
  particularly interesting situation related to pedestrian dynamics
  and swarming intelligence. The present work addresses these three
  issues while taking into account permanent and transient marks left
  in the visited nodes. Different topologies are considered for trail
  reconstruction and trail source identification, including four
  complex networks models and four real networks, namely the Internet,
  the US airlines network, an email network and the scientific
  collaboration network of complex network researchers. Our results
  show that the topology of the network influence in trail reconstruction,
  source identification and agent dynamics.

\end{abstract}


\maketitle

\emph{`... when you have eliminated the impossible, whatever remains,
  however improbable, must be the truth.'} (Sir A. C. Doyle, Sherlock
Holmes) \vspace{0.5cm}

\section{Introduction}

Complex networks have become one of the leading paradigms in science
thanks to their ability to represent and model highly intricate
structures (e.g., \cite{Albert_Barab:2002, Newman03:SIAM,
Boccaletti:2006, Costa_surv:2007}). However, as a growing number of
works have shown (e.g., \cite{Newman03:SIAM, Boccaletti:2006}) the
dynamics of systems whose connectivity is defined by complex networks
is often even more complex and interesting than the connectivity of
the networks themselves. One particularly interesting type of
non-linear dynamics involves the evolution of \emph{trails} left by
moving agents during random walks or dilation processes along the
network . The term ``dilation'' refers to the progressive visiting of
neighboring nodes after starting from one or more nodes.  For
instance, starting from node $i$, at each subsequent time the
neighbors of $i$ are visited, then their unvisited neighbor, and so
on, defining a hierarchical system of neighborhoods
(e.g.~\cite{Faloutsos99, Costa_PRL:2004, Costa_JSP:2006}). Although
the dynamics is being described as agents visiting network sites, it
can be considered also as the evolution on activity in the nodes of
the network, where each network edge represents the possibility of
activity propagation between the respective nodes. Another important
related problem involves attempts to recover incomplete trails. In
other words, in cases in which only partial evidence is available to
observation, it becomes important to try to infer the full set of
visited nodes.

The emergencey of trails has been studied as representing an
interesting type of self-organizational system. Helbing et
al.~\cite{Helbing97:Nature} proposed a model of pedestrian motion in
order to explore the evolution of trails in urban green areas. Also,
trails have been considered in swarming intelligence
analysis~\cite{Bonabeau1999sin,kennedy2001si} not only as a means to
understand animal behavior~\cite{Helbing97:PRE}, but also as a source
of insights for new optimization and routing algorithms~\cite{Dorigo1997act,dorigo2004aco}. 
These works considered the evolution of trails in regular grids. However, the
communication structures where the trail can be defined are not
homogeneous in many cases. Many systems, such as the
Internet~\cite{Faloutsos99}, social relationships~\cite{Newman03:PRE},
the distribution of streets in cities~\cite{Rosvall05:PRL} and the
connections between airports~\cite{Guimera04:EPJB} are defined by a
irregular topology --- more specifically, most of these systems are
represented by scale-free networks~\cite{Costa_surv:2007}. Here, we
study the influence of different topologies in trails recovery, source
identification and agent dynamics.

The analysis of trails left in complex networks can have many useful
applications. For instance, in information networks the recovery of
the trail left by a spreading virus on the Internet can be useful to
identify the source of contamination and propose strategies for
computer immunization. Similarly, the identification of the origin of
rumors, diseases, fads and opinion formation~\cite{Rodrigues05:IJMPC}
are important to understand the human communication dynamics. Another
relevant problem is related to traffic improvement and security. In
the former case, identification of the covered trails by packages
exchanged between computers can help the development of optimal
routing paths. In the latter, the source of terrorism strategies and
drug trafficking can be determined by analysis of clues identified in
social and airline networks. The analysis of trails can also have
useful applications in biology. For instance, in ecology, trails
analysis can be applied to quantify the interference of human activity
in animal behavior and to identify focus of pollution. In
paleontology, the recovery of the trails of animal displacement by
fossil analysis can help the understanding of diversification between
species. In epidemiology, the identification of disease source can
help to stop the spreading process as well as to devise effective
prevention strategies.

In order to properly represent trails occurring in complex networks,
we associate state variables to each node $i$, $i=1, 2, \ldots, N$, of
the network.  The trail is then defined by marking such variables
along the respective dynamical process. Only trails generated by
self-avoiding random walks and dilations are considered in the current
work, which are characterized by the fact that a node is never visited
more than once. We restrict our attention to binary trails,
characterized by binary state variables~\footnote{In other words, a
node can be marked as either already visited (1) or not (0).  Graded
states e.g., indicating the time of the visit, are considered only on
analysis of dynamical agents propagation in Section~\ref{Sec:agents}.}. The
types of trails can be further classified by considering the marks to
be permanent or transient. In the latter case, the mark associated to
a node can be deleted after the visit.  While many different transient
dynamics are possible, we restrict our attention to the following two
types: (i)
\emph{Poissonian}, where each mark has a fixed probability of being
removed after the visit; and (ii) \emph{Evanescent}, where the only
observable portion of the trail correspond to the node(s) being
currently visited.

The current work addresses the problem of recovering trails in complex
networks and identifying their origin, while considering permanent and
transient binary marks in four different networks models, namely
Erd\H{o}s-R\'enyi, Watts-Strogatz, Barab\'asi-Albert, and
Dorogovtsev-Mendes-Samukhin models; and four real networks: the
Internet at the Autonomous System level, the US airlines network, an
email network from the University Rovira i Virgili and the scientific
collaboration network of complex network researchers. We also consider
the analysis of agents propagation considering the four networks
models. The next sections start by presenting the basic concepts in
complex networks and trails and follow by reporting the simulation
results, with respective discussions.

\section{Basic Concepts in Complex Networks and Trails}

An undirected complex network (or graph) $G$ is defined as $G=(V,Q)$, where $V$ is the set of $N$ nodes and
$Q$ is the set of $E$ edges of the type $\{i,j\}$, indicating that nodes $i$ and $j$ are bidirectionally
connected.  Such a network can be completely represented in terms of its \emph{adjacency matrix} $K$, such
that the presence of the edge $\{i,j\}$ is indicated as $K(i,j)=K(j,i)=1$ [otherwise $K(i,j)=K(j,i)=0$].  The
degree of a node $i$ corresponds to the number of edges connected to it, which can be calculated as
$k(i)=\sum_{j=1}^{N}K(i,j)$. The clustering coefficient is related to the presence of triangles (cycles of
length three) in the network~\cite{Watts:1998}.  The clustering coefficient of a node $i$ is given by the
ratio between the number of edges among the neighbors of $i$ and the maximum possible number of edges among
these neighbors; the clustering coefficient of the network is the average of the clustering coefficient of
its nodes.

This article considers four theoretical network models and four real complex networks. The network models are
(a) Erd\H{o}s-R\'enyi --- ER~\cite{ErdosRenyi:1961}, (b) Watts-Strogatz --- WS~\cite{Watts:1998}, (c)
Barab\'asi-Albert --- BA~\cite{Albert_Barab:2002} and (d) Dorogovtsev-Mendes-Samukhin ---
DMS~\cite{Dorogovtsev:PRL00}. In the first model, networks are constructed by considering constant
probability $\lambda$ of connection between any pair of nodes; in the second, networks start with a regular
topology, whose nodes are connected in a ring to a defined number $\kappa$ of neighbors in each direction,
and later the edges are rewired with a fixed probability; networks of the third and fourth models are grown
by starting with $m_0$ nodes and progressively adding new nodes with $m$ edges, which are connected to the
existing nodes with probability proportional to their degrees (e.g., \cite{Albert_Barab:2002}). The DMS model
differs from the BA model by adding an initial attractiveness $k_0$ to each node, independent of its degree.
When $k_0 = 0$, the DMS model is similar to the BA model~\cite{Dorogovtsev:PRL00}.  All simulations
considered in this work assume that the networks have the same number of nodes $N=1000$ and average degree
$\left< k \right > = 2m = \lambda (N-1) = 2\kappa = 4$.  The real networks considered in this work are the
Internet at the level of autonomous systems \footnote{The considered data in our work is available at the web
site of the National Laboratory of Applied Network Research (\texttt{http://www.nlanr.net}). We used the data
collected in Feb.  $1998$.}, the US Airlines~\cite{pajek-data}, the e-mail network from the University Rovira
i Virgili (Tarragona)~\cite{guimera2003ssc} and the scientific collaboration of complex networks
researchers~\footnote{The scientific collaboration of complex networks researchers was compiled by Mark
Newman from the bibliographies of two review articles on networks (by Newman~\cite{Newman03:SIAM} and
Boccaletti \emph{et al.}~\cite{Boccaletti:2006}).}.

\begin{figure}
 \begin{center}
   \includegraphics[width=0.7\columnwidth]{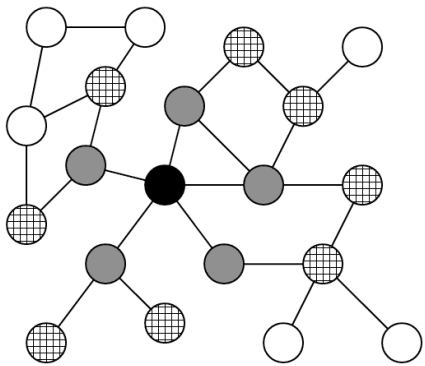}
   \caption{Dilating trail with two levels in a simple network.  The
     origin of this two-hierarchy-trail is the black node, whose
     immediated neighbors are marked in gray.  The nodes with the
     crossed pattern correspond to the neighbors of the neighbors of
     the source of the trail.  The respective evanescent trail would
     include only the crossed nodes.  A Poissonian version of this
     trail would imply a ratio $\gamma$ of unmarked (and unobservable)
     nodes.}
   \label{fig:dils}
\end{center}
\end{figure}

\emph{Trails} are generated while subsets of the nodes $V$ are visited during the evolution of random walks
or dilations through the network. We assume that just one trail is allowed at any time in a complex network.
We consider only self-avoiding random walks, in which no node is visited more than once.  At each node, the
agent chooses a new node to be visited at random among the not yet visited neighbors of the node.  To
understand the dilation process, consider $\nu(i)$ the set of neighbors of node $i$.  Starting with $i_o$,
the initial node of the propagation (origin), all nodes in $\nu(i_0)$ are visited; after that, for all
$j\in\nu(i_0)$, the nodes in $\nu(j)$ not yet visited are recursively visited; this process is repeated for a
given number of neighborhood hierarchies (e.g., \cite{Faloutsos99, Costa_PRL:2004, Costa_EPJB:2006,
Costa_JSP:2006}); see Fig.~\ref{fig:dils}.

\begin{figure}
 \begin{center}
   \includegraphics[width=0.7\columnwidth]{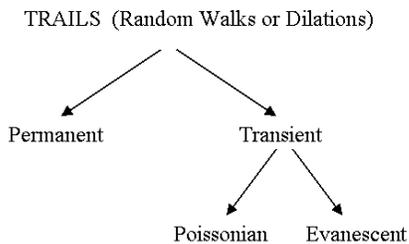}
   \caption{Trails, including those defined by random walks and
     dilations, can be subdivided as being permanent or transient.
     The latter type can be further subdivided into Poissonian and
     Evanescent.}
   \label{fig:class}
\end{center}
\end{figure}

In order to represent trails, we associate two binary state variables $v(i)$ and $s(i)$ to each node $i$,
which can take the values 0 (not yet visited) or 1 (visited). The state variables $v(i)$ indicate the real
visits to each node but are available only to the moving agents, the state variables $s(i)$ are the ``marks''
of the visits yet available for observation, providing not necessarily complete information about the visits.
The structure of the network is assumed to be known to the observer and possibly also to the moving agent(s).
Such a situation corresponds to many real problems. For instance, in case the trail is being defined as an
exploring agent moves through unknown territory, the agent may keep some visited places marked with physical
signs (e.g., flags or stones) which are accessible to observers, while keeping a complete map of visited
sites available only to her/himself.  Trails are here classified as \emph{permanent} or \emph{transient}. In
the case of permanent trails, $s(i)=v(i)$, i.e.\ all visited nodes are known.  In the transient type, the
state variables $s(i)$ of each node $i$ can be reset to zero after being visited.  Transient trails can be
further subdivided into: (i)~\emph{Poissonian}, characterized by the fact that each visited node has a fixed
probability $\gamma$ of not being observed, i.e.\ for nodes with $v(i)=1$, $s(i)$ is $1$ with probability
$1-\gamma$ and $0$ with probability $\gamma$ (nodes with $v(i)=0$ always have $s(i)=0$); and
(ii)~\emph{Evanescent}, where only the last visited nodes are accessible to the observer.
Figure~\ref{fig:class} shows a classification of the main types of trails considered in this work.

The \emph{real extension} of a trail is defined as being equal to the sum of the state variables $v(i)$. The
\emph{observable extension} of a trail is equal to the sum of the state variables $s(i)$. Given a trail, we
can define the \emph{observation error} as being equal to
\begin{equation}
  \epsilon = \sum_{i=1}^{N} \left[ 1-\delta(v(i),s(i))\right],
\end{equation}
where $\delta(a,b)$ is the Kronecker delta function, yielding one in case $a=b$ and zero otherwise. Note that
this error measures the incompleteness of the information provided to the observer.  It is also possible to
normalize this error by dividing it by $N$, so that $0 \leq \epsilon \leq 1$; this normalization is not used
in this work.

It is assumed that the observer will try to recover the original, complete, trail from its observation.  In
this case, the observer applies some heuristic in order to obtain a recovered trail specified by an
additional set of state variables $r(i)$ ($r(i) = 1$ if node $i$ is in the recovered trail).  Such a
heuristic may take into account the \emph{overlap error} between the observable states $s(i)$ and the
recovered values $r(i)$, defined as
\begin{equation}
  \xi = \sum_{i=1}^{N} \left[ 1-\delta(s(i),r(i))\right].
\end{equation}
Note that as the observer has no access to $v(i)$, the recovery error has to be estimated using $s(i)$.  The
actual \emph{recovery error}, which can be used to infer the quality of the recovery, is given by
\begin{equation}
  \rho = \sum_{i=1}^{N} \left[ 1-\delta(v(i),r(i))\right].
  \label{recovery_error}
\end{equation}

\begin{figure}
 \begin{center}
   \includegraphics[width=0.7\columnwidth]{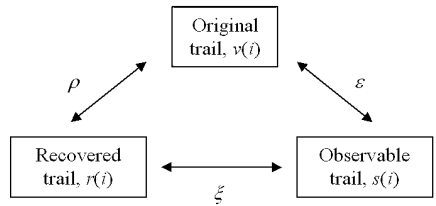}
   \caption{The three state variables associated with each network
     node $i$ and the defined errors $\epsilon$, $\xi$ and $\rho$.}
   \label{fig:errs}
\end{center}
\end{figure}

Figure~\ref{fig:errs} illustrates the three state variables related to each network node and the respectively
defined errors.

When using recovery heuristics based on the evaluation of the overlap error, it may happen that two or more
different recovered trails yield the same overlap error.  In this case, it is interesting to consider two
additional parameters in order to quantify the quality of the recovery: (i)~the number $M$ of estimated
trajectories corresponding to the minimum overlap error; and (ii)~the fraction $f$ of times that the correct
source can be found among the $M$ recovered trails.  When average values of $M$ and $f$ are close to $1$, it
means that the recovery strategy is precise.

\section{Considered Problems}

Although the problem of trail analysis in complex networks is potentially very rich and can be extended to
many possible interesting situations, for simplicity's sake we restrict our interest to the three following
cases:
\begin{description}

\item[Poissonian trails from random walks:] Because the consideration
  of permanent and evanescent trails left by random walks are
  trivial~\footnote{Permanent trails left by random walks requires no
    recover, while their source should necessarily correspond to any
    of its two extremities.  Evanescent trails defined by random walks
    are meaningless, as only the current position of the single agent
    is available to the observer.}, we concentrate our attention on
  the problem of recovering Poissonian trails left by single moving
  agents during random walks.  Once such a trail is recovered, its
  source can be estimated as corresponding to one of its two
  extremities; we do not consider the problem of source identification
  for this kind of trail.  The recovery error is used to measure the
  quality of the reconstructed trail.

\item[Poissonian trails from dilations:] In this case, only a fraction
  of the nodes visited by the dilating process is available to the
  observer.  Two problems are of interest here, namely recovering the
  trail and identifying its origin.  To quantify the quality of the
  recovery, we evaluate the average values of the number of trails
  with minimal overlap error $\langle M\rangle$ and the fraction of
  correct source identifications $\langle f\rangle$.

\item[Evanescent trails from dilations:] In this type of problem, only
  the currently visited nodes are available to the observer, which is
  requested to reconstruct the trail and infer its possible origin.
  This corresponds to the potentially most challenging of the
  considered situations.  Note that this case too is subject to random
  removal of marks, i.e.\ the values of $s(i)$ are not only of the
  evanescent type but subjected to be randomly changed to $0$.  The
  results are evaluated by computing $\langle M\rangle$ and $\langle
  f\rangle$.

\end{description}

\section{Strategies for Recovery and Source Identification}

Several heuristics can be possibly used for recovering a trail from the information provided by $K$ and
$s(i)$.  In this work, we consider a strategy based on the topological proximity on the network between nodes
with $s(i)=1$ that are not connected.  In the case of trails left by random walks, the following algorithm is
used:
\begin{enumerate}

\item Initialize a list $r$ as being equal to $s$;

\item For each node $i$ with $s(i)=1$:
  \begin{enumerate}
  \item identify the node $j$ with $r(i)=1$ which is connected to at
    most one other node with $r(i)=1$ and is closest to $i$ (in the
    sense of shortest topological path, but excluding shortest paths with
    length 0 or 1 in the network);
  \item obtain the list $L$ of nodes linking $i$ to $j$ through the
    respective shortest path (if more than one shortest path exist,
    one of them is chosen at random);
  \item for each node $k$ in $L$, make $r(k)=1$.
  \end{enumerate}

\end{enumerate}

After all nodes with $s(i)=1$ have been considered, the recovered trail will be given by the nodes with
$r(i)=1$.

\begin{figure}
 \begin{center}
   \includegraphics[width=0.7\columnwidth]{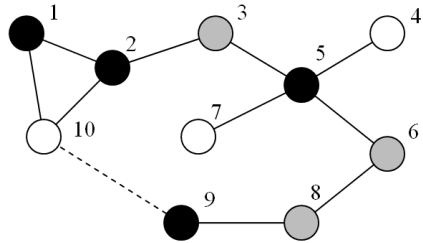}
   \caption{Example of simple Poissonian trail in a network.  The
     black nodes correspond to $s$, the original trail included the
     black and gray nodes.}
   \label{fig:ex_rw}
\end{center}
\end{figure}

Figure~\ref{fig:ex_rw} illustrates a simple Poissonian random walk trail, where the black nodes are those in
$s$.  The original trail is composed of the nodes in $s$ plus the gray nodes.  It can be easily verified that
the application of the above reconstruction heuristic will properly recover the original trail in this
particular case. More specifically, we would have the following sequence of operations:
\begin{itemize}

\item[\bf Step 1:] node 1 connected to node 5 through the shortest
  path $(1, 2, 3, 5)$;

\item[\bf Step 2:] node 2 connected to node 5 (no effect);

\item[\bf Step 3:] node 5 connected to node 2 (no effect);

\item[\bf Step 4:] node 9 connected to node 5 through the shortest
  path $(9, 8, 6, 5)$.

\end{itemize}

However, if the dashed edge connecting nodes 9 and 10 were included into the network, a large recvery error
would have been obtained because the algorithm would link node 9 to node 1 or 2 and not to node 5.

A different strategy is used for recovery and source identification in the case of dilation trails, which
involves repeating the dilation dynamics while starting from each of the network nodes.  The most likely
recovered trails are those corresponding to the smallest obtained overlap error.  Note that more than one
trail may correspond to the smallest error. Also, observe that the possible trail sources are simultaneously
determined by this algorithm.  Actually, it is an interesting fact that complete recovery of the trail is
automatically guaranteed once the original source is properly identified.  This is an immediate consequence
of the fact that the recovery strategy involves the reproduction of the original dilation, so that the
original and obtained trails for the correct source will necessarily be identical.

\begin{figure}
 \begin{center}
   \includegraphics[width=0.7\columnwidth]{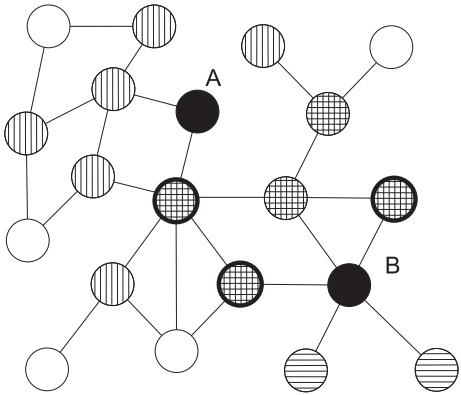}
   \caption{Simple illustration of the source of ambiguity in trail
     source determination. See text for explanation.}
   \label{fig:amb_dils}
\end{center}
\end{figure}

Some additional remarks are required in order to clarify the reason why more than one trail can be identified
as corresponding to the minimal overlap error in Poissonian dilation trails. Figure~\ref{fig:amb_dils}
illustrates a simple network with two trails extending through two hierarchies, one starting from the source
A and the other from B, which are respectively identified by the vertical and horizontal patterns. Note that
some of the nodes are covered by both trails, being therefore represented by the crossed pattern. Now, assume
that the original trail was left by A but that the respectively Poissonian version only incorporated the
three nodes with thick border (i.e.\ all the other nodes along this trail were deleted before presentation to
the observer).  Because the three nodes are shared by both trails, the same overlap error will be obtained by
starting at nodes A or B. It is expected that the higher the value of $\gamma$, the more ambiguous the source
identification becomes.

When many possible recovered trails with the same overlap error are found, i.e.\ when $M>1$, the
identification of the source is ambiguous.  To take this fact into account, in that cases we consider that
each of the possible sources is as good as the other, and therefore can be used as the evaluated source;
therefore we make $f=1/M$.

\section{Simulation Results and Discussion}

To evaluate the recovery strategies under different topologies, randomly generated trails are studied in the
ER, SW, BA, and DMS network models and the networks of Internet (AS), US Airlines, e-mail and scientific
collaboration, as indicated previously. The following sections present e discuss those results.

\subsection{Network models}

Each considered network model is formed by $N = 1\,000$ nodes and average degree $\left< k \right> = 4$.  All
random walk trails were Poissonian with real extent equal to 20 nodes and $\gamma=0.1, 0.2, \ldots, 0.8$. All
dilation trails took place along 2 hierarchies, while the respective Poissonian and evanescent cases assumed
$\gamma=0.1, 0.2, \ldots, 0.8$. In order to provide statistically significant results, each configuration
(i.e.\ type of network, trail and $\gamma$) was simulated 100~times. The rewiring probability in WS model is
the same as in ER model, i.e.\ $p = \left< k \right>/(N-1)$. The initial connectivity in DMS networks models
is $k_0 = 5$.

\begin{figure}
  \begin{center}
    \subfigure[]{\includegraphics[width=0.45\columnwidth]{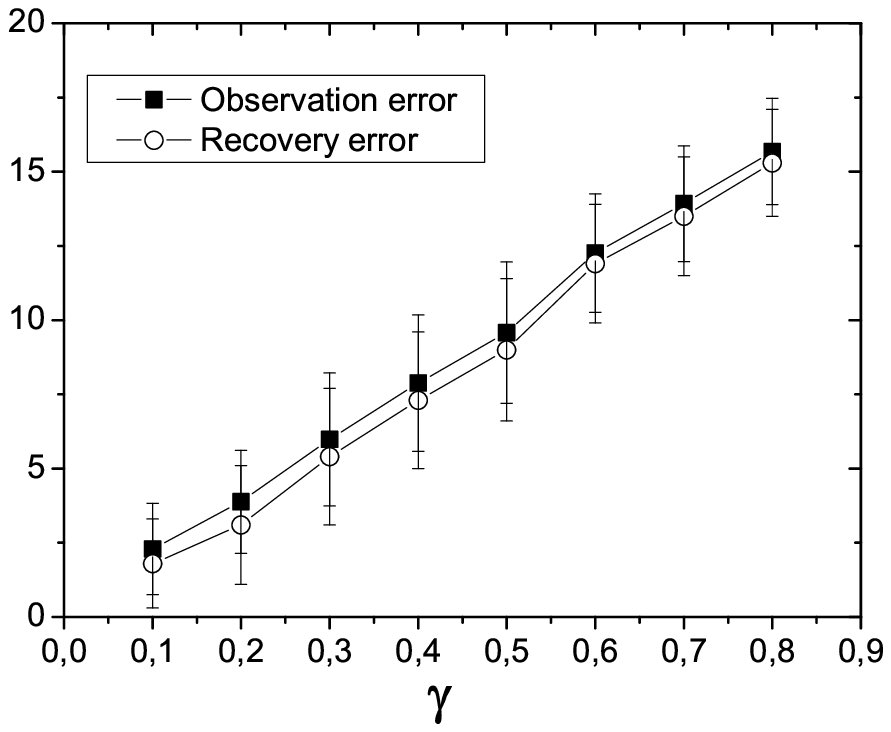}}
    \subfigure[]{\includegraphics[width=0.45\columnwidth]{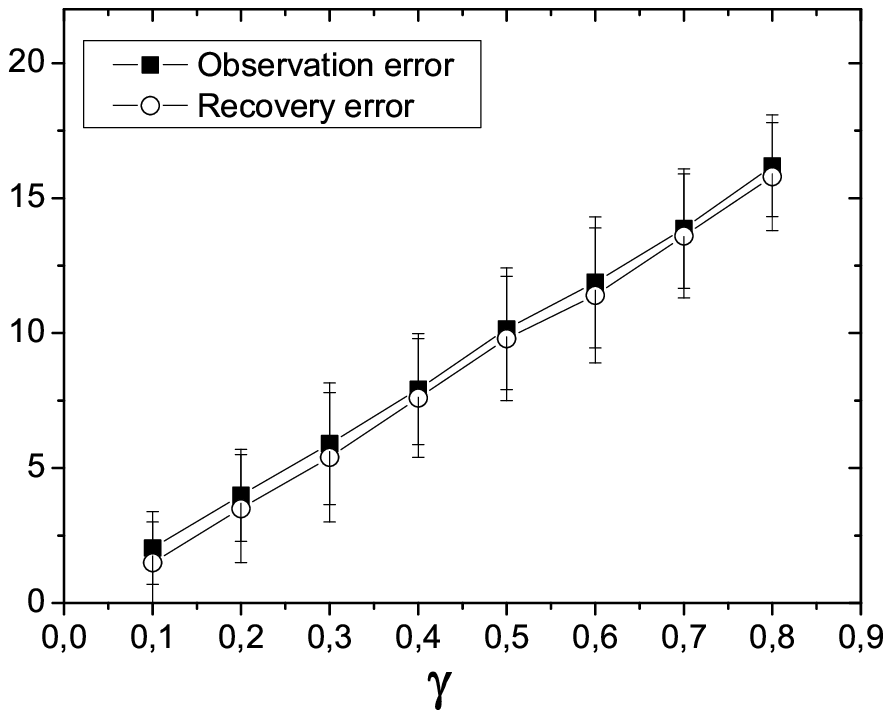}}
    \subfigure[]{\includegraphics[width=0.45\columnwidth]{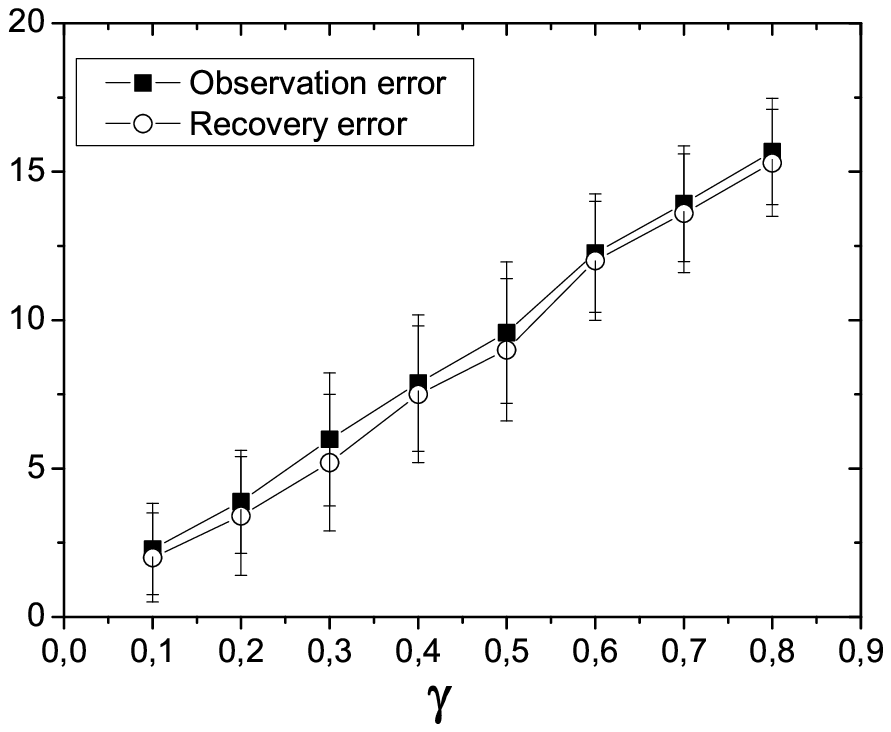}}
    \subfigure[]{\includegraphics[width=0.45\columnwidth]{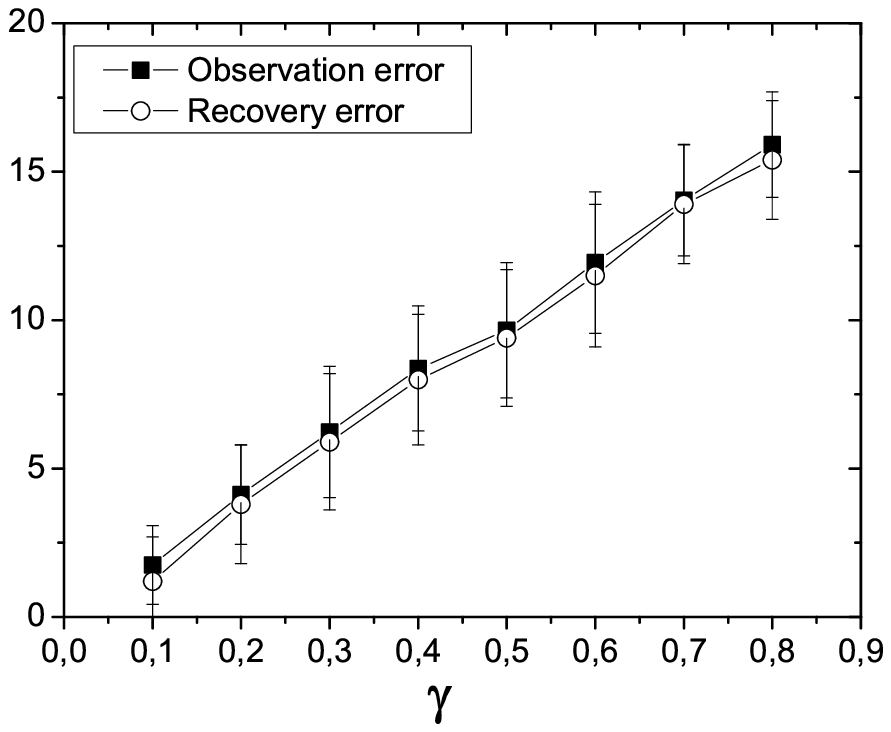}}
  \end{center}
  \caption{The observation error (black squares) and the recovery
    errors (white circles) obtained by using the recovery algorithm
    for for Poissonian trails from random walks in the (a)~ER, (b)~SW,
    (c)~BA and (d)~DMS network models.}
  \label{fig:recmodels}
\end{figure}

Figure~\ref{fig:recmodels} shows the average observation and recovery errors, with respective standard
deviations, obtained for the Poissonian random walk trails in the four considered network models. The figure
indicates an almost linear increase of the recovery error with $\gamma$. Such a monotonic increase is
explained by the fact that the higher the value of $\gamma$, the more incomplete the observable states
become. As the recovery of trails with more gaps will necessarily imply more wrongly recovered patches, the
respective error therefore will increase with $\gamma$. Also, as can be seen by a comparison between
observation and recovery errors, the adopted recovery heuristic allowed moderate results for all considered
network models, without significative differences among the models, which suggests that such a recovery
strategy is independent of the network topology.

\begin{figure}
  \begin{center}
    \subfigure[]{\includegraphics[width=0.45\columnwidth]{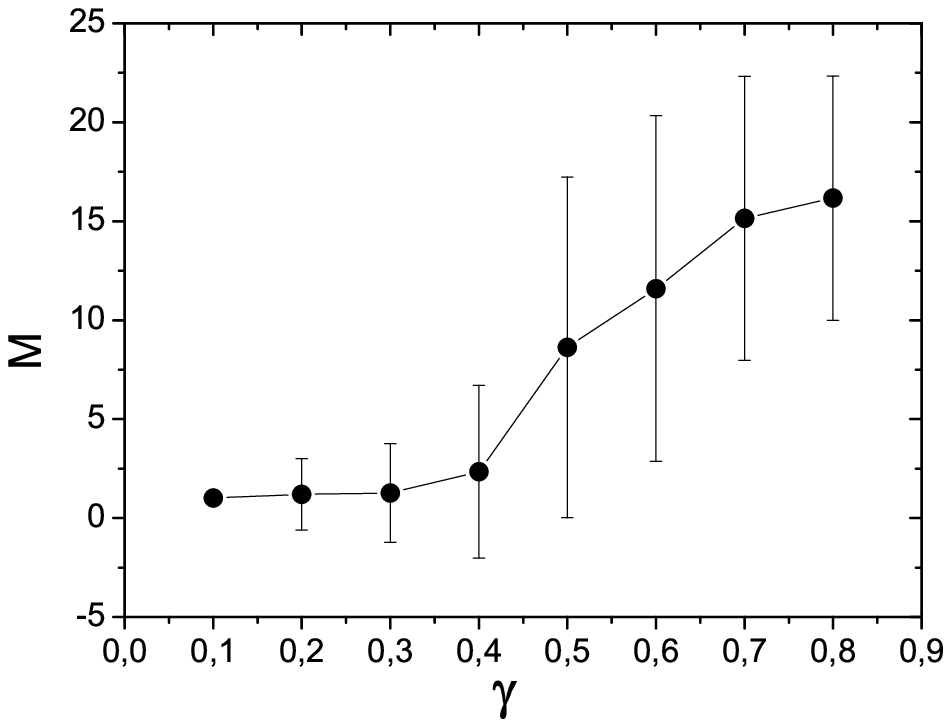}}
    \subfigure[]{\includegraphics[width=0.45\columnwidth]{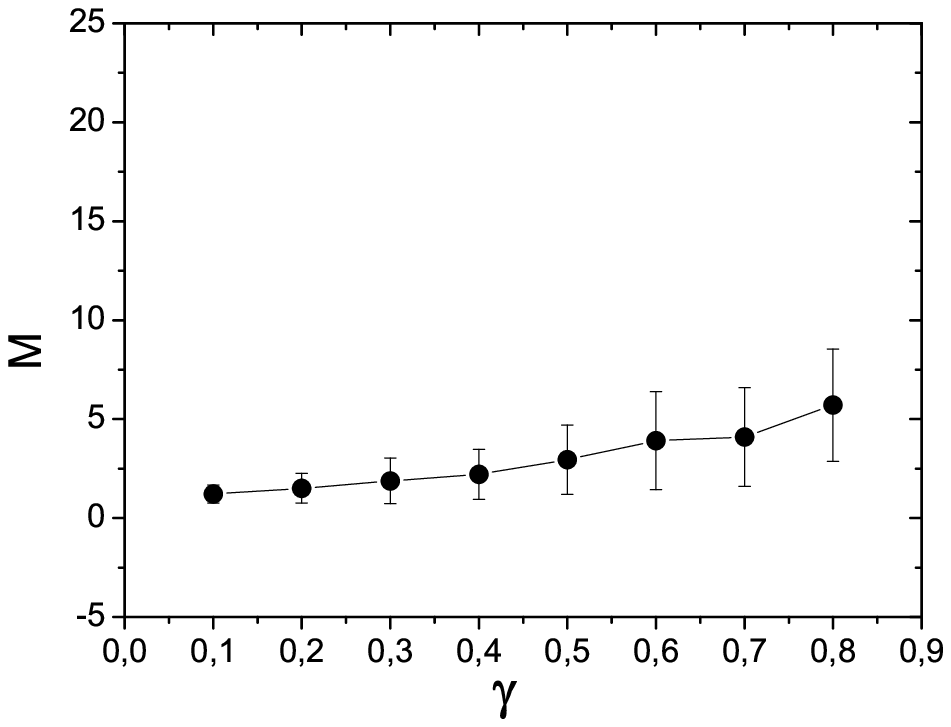}}
    \subfigure[]{\includegraphics[width=0.45\columnwidth]{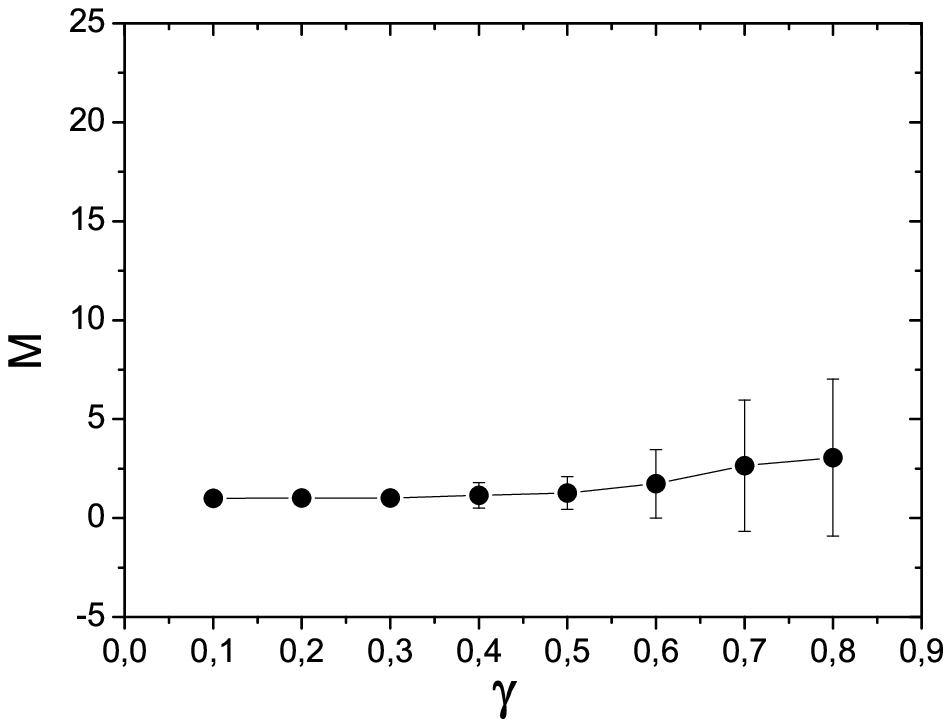}}
    \subfigure[]{\includegraphics[width=0.45\columnwidth]{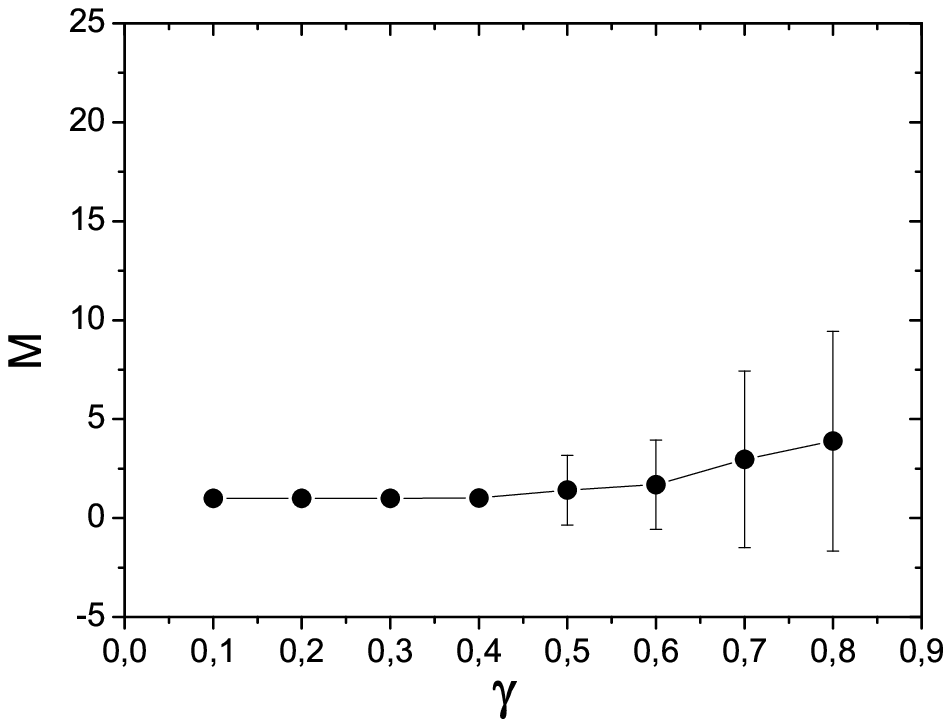}}
  \end{center}
  \caption{The average and standard deviations, in terms of $\gamma$,
    of the number $M$ of detected trails corresponding to the minimal
    overlap error with respect to Poissonian dilation trails obtained
    for ER~(a), SW~(b), BA~(c) and DMS~(d) network models.}
  \label{fig:Mmodels}
\end{figure}

Figure~\ref{fig:Mmodels} gives the average and standard deviation of $M$ for Poissonian dilation trails
corresponding to the minimal overlap error $\xi$ for ER, SW, BA and DMS networks.  In all of these models,
the average and standard deviation values of $M$ tend to increase with $\gamma$, starting at $\left<M
\right>=1$.  This effect is a consequence of the fact that, the more sparse the information about the real
trail, the more likely it is to cover the observable states $s$ with dilations starting from different nodes.
Interestingly, the increase of $\left<M \right>$ is substantially more accentuated for ER networks, and BA
networks are the least subject to source determination ambiguities.

\begin{figure}
  \begin{center}
    \subfigure[]{\includegraphics[width=0.45\columnwidth]{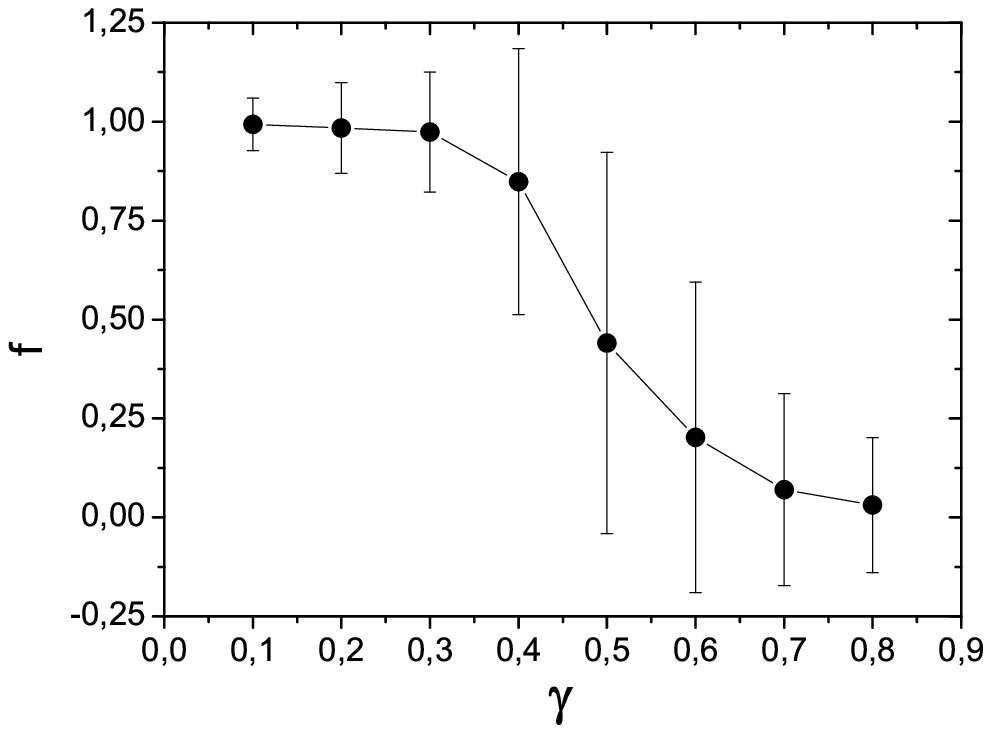}}
    \subfigure[]{\includegraphics[width=0.45\columnwidth]{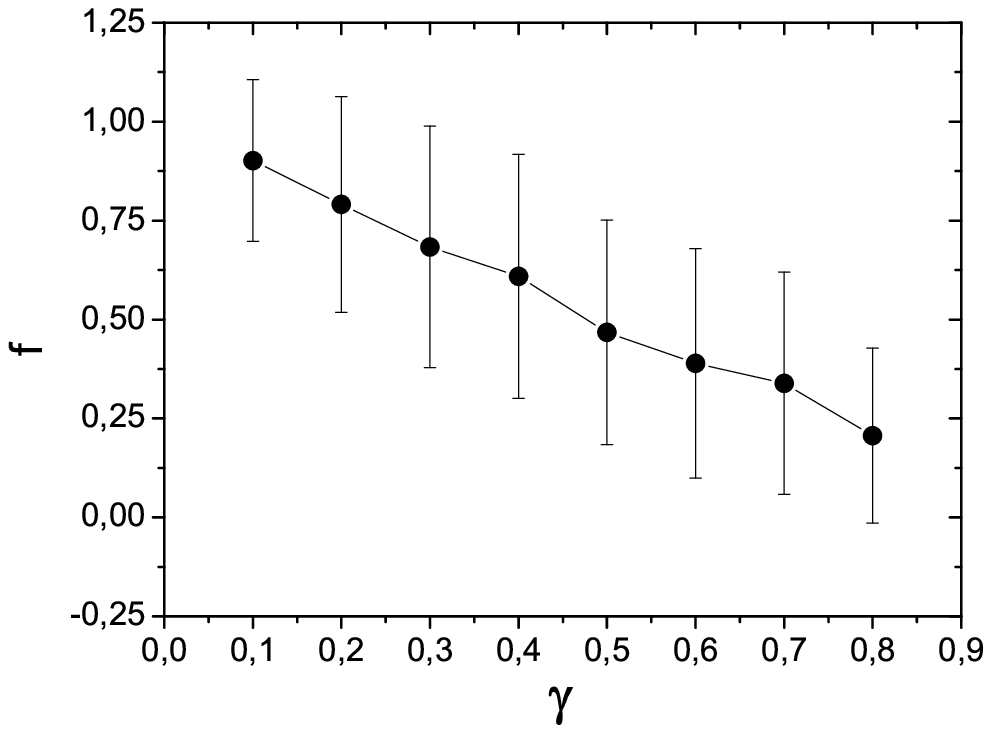}}
    \subfigure[]{\includegraphics[width=0.45\columnwidth]{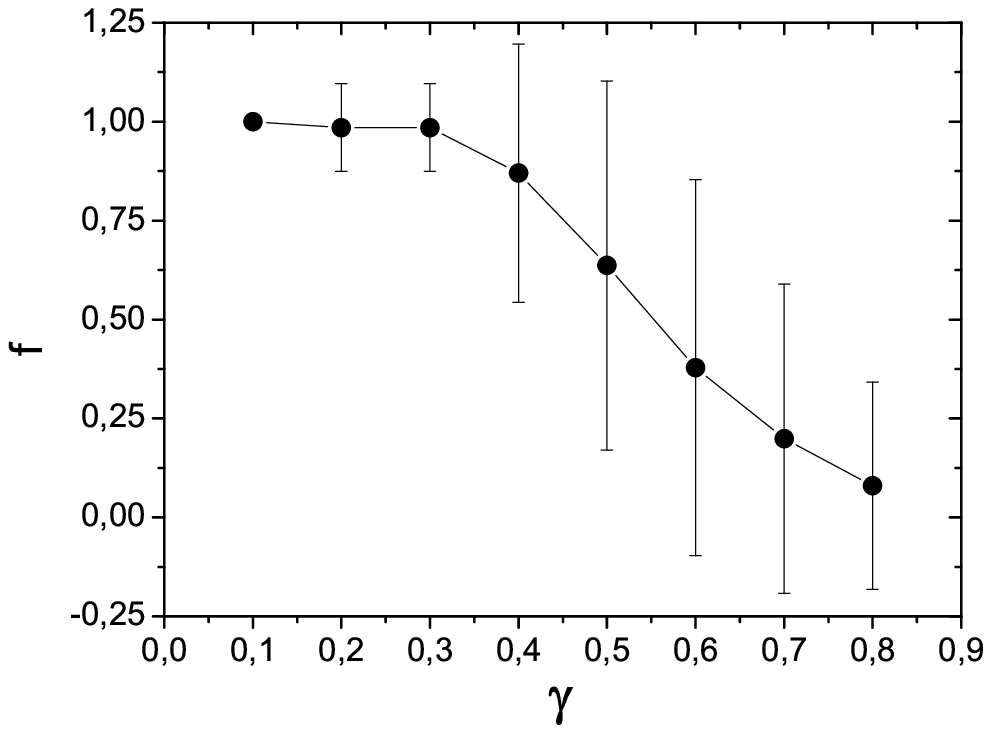}}
    \subfigure[]{\includegraphics[width=0.45\columnwidth]{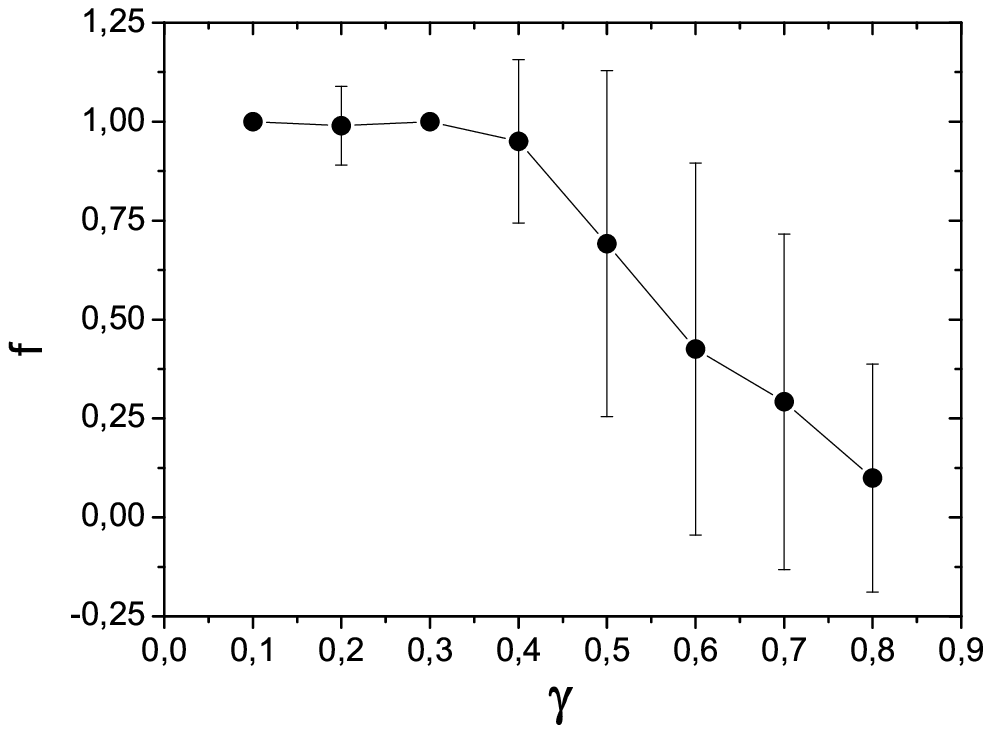}}
  \end{center}
  \caption{The average (and standard deviations) of the flag $f$
    indicating that the correct source has been identified among the
    detected trails with minimal overlap error $\xi$ in the recovery
    of Poissonian dilation trails for ER~(a), SW~(b), BA~(c) and
    DMS~(d) network models.}
  \label{fig:fmodels}
\end{figure}

For the Poissonian dilation trails, the average $\left< f \right>$ (and standard deviation) of the flag $f$
is given in terms of $\gamma$ in Figure~\ref{fig:fmodels} for ER, SW, BA, and DMS networks.  It is clear from
these results that the average number of times, along the realizations, in which the correct source is
identified among those trails corresponding to the minimal overlap error $\xi$ tends to decrease with
$\gamma$. This is a direct consequence of the fact that higher values of $\gamma$ imply substantial
distortions to the original trail, ultimately leading to shifts in the identification of the correct source.
The behavior of $\left< f \right>$ is similar for ER, BA and DMS network models, with a sharp decrease for
$\gamma \gtrsim 0.3$. For SW networks, on the other hand, $\left< f \right>$ has a smooth decrease. The
sources of the trails are best identified for ER, BA and DMS when $\gamma \lesssim 0.3$.  For higher values
of $\gamma$, the sources are best identified for SW network models.

\begin{figure}
  \begin{center}
    \subfigure[]{\includegraphics[width=0.45\columnwidth]{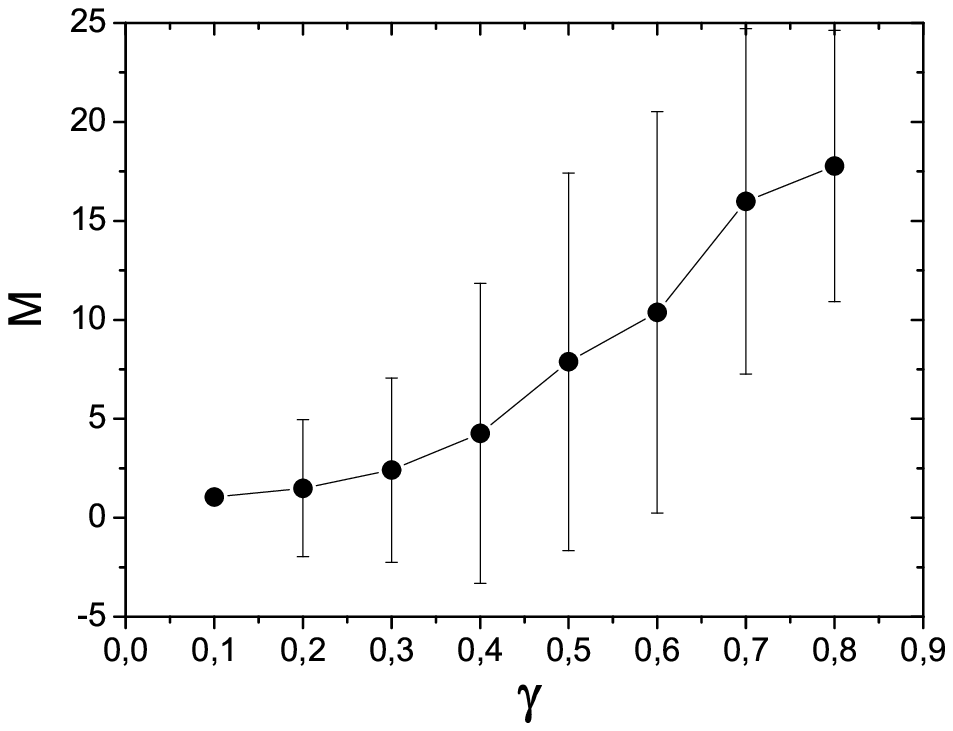}}
    \subfigure[]{\includegraphics[width=0.45\columnwidth]{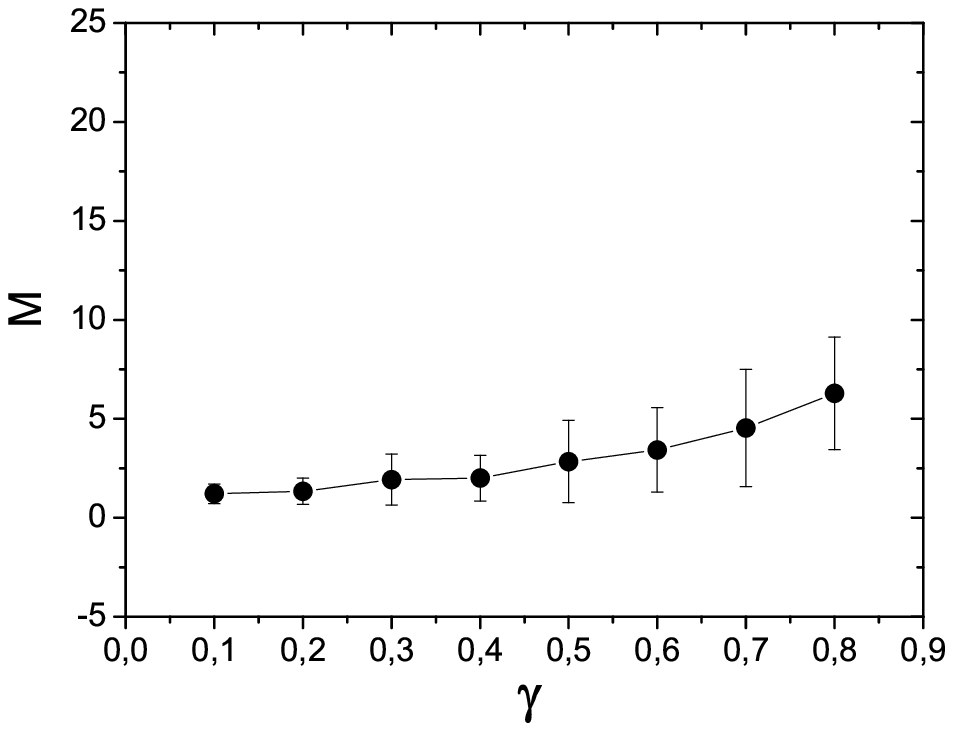}}
    \subfigure[]{\includegraphics[width=0.45\columnwidth]{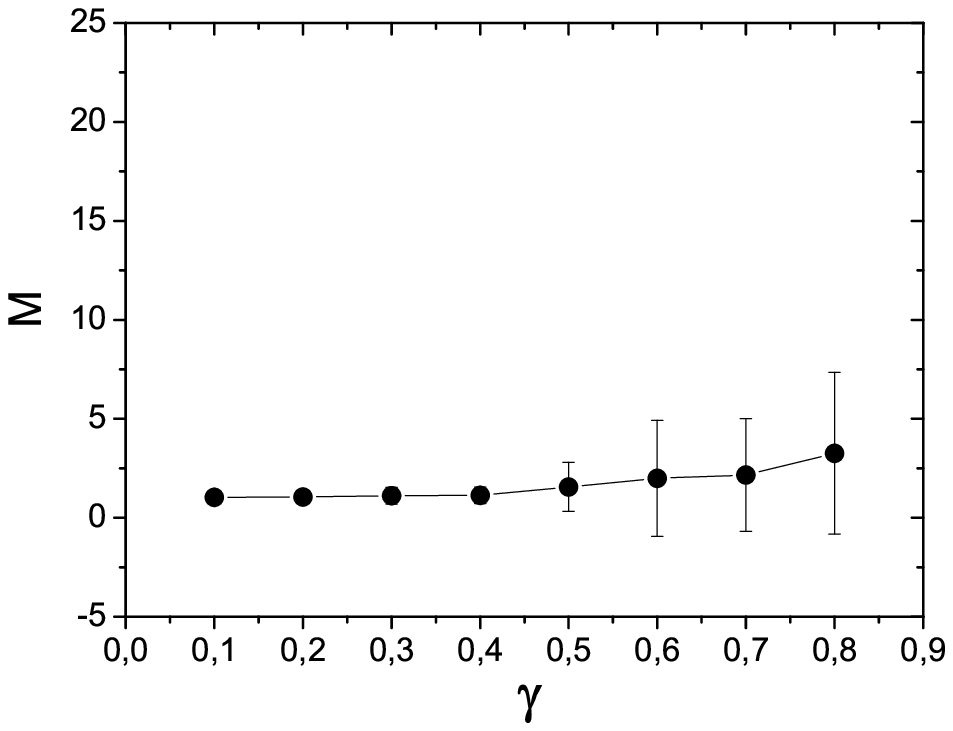}}
    \subfigure[]{\includegraphics[width=0.45\columnwidth]{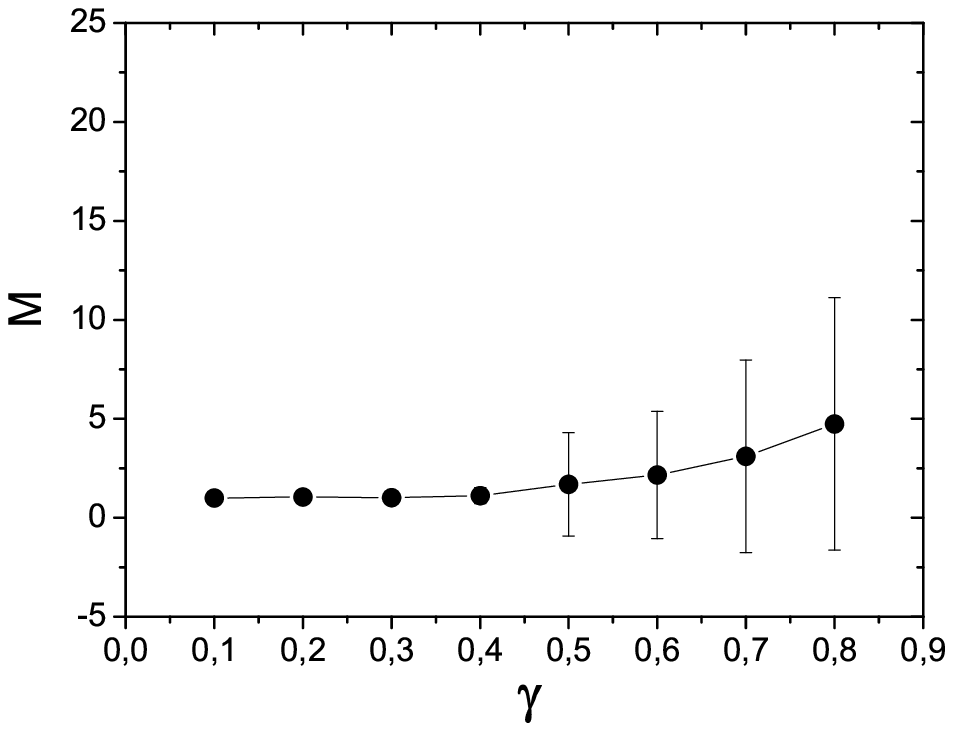}}
  \end{center}
  \caption{The average and standard deviations, in terms of $\gamma$,
    of the number $M$ of detected evanescent trails corresponding to
    the minimal overlap error obtained for ER~(a), SW~(b), BA~(c) and
    DMS~(d) network models.}
  \label{fig:mevamodels}
\end{figure}

Finally, we turn our attention to transient dilation trails of the \emph{evanescent} category.  Recall that
in this type of trails only the current position of the trail (i.e.\ its border) is available to the
observer.  Figure~\ref{fig:mevamodels} presents the average and standard deviation of $M$ obtained, in terms
of $\gamma$, for the ER, SW, BA and DMS network models.  The result is similar to the case of Poissonian
trails (Fig.~\ref{fig:Mmodels}), with the recovery strategy having the worst results for ER networks, and
similar results among the other models.  But for the evanescent trails $M$ grows more gradually than for
Poissonian trails.

\begin{figure}
  \begin{center}
    \subfigure[]{\includegraphics[width=0.45\columnwidth]{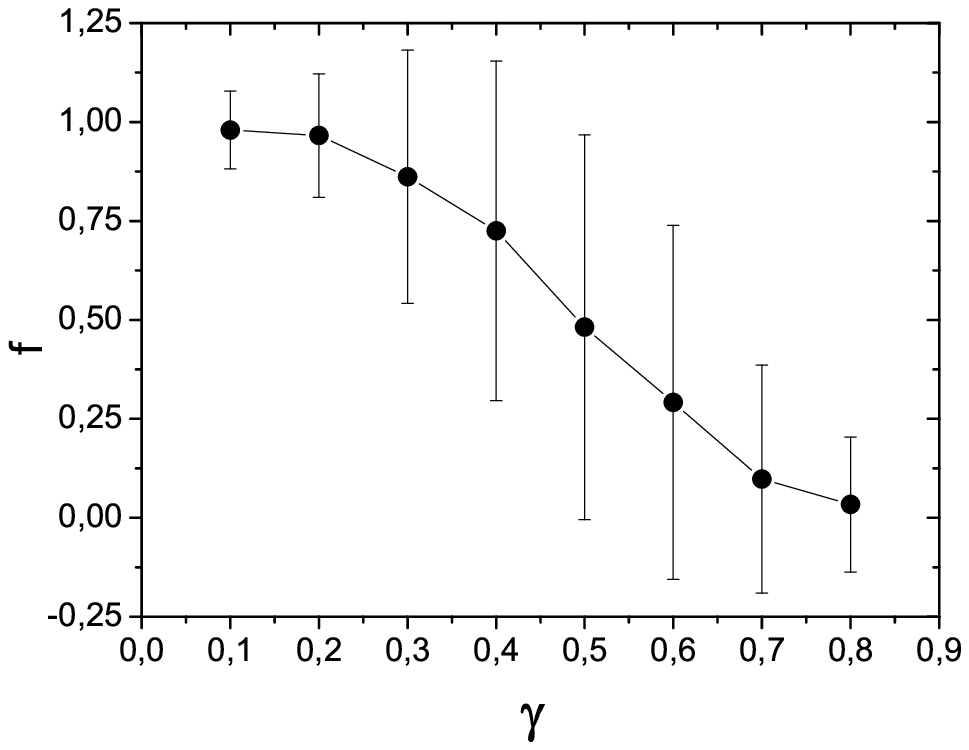}}
    \subfigure[]{\includegraphics[width=0.45\columnwidth]{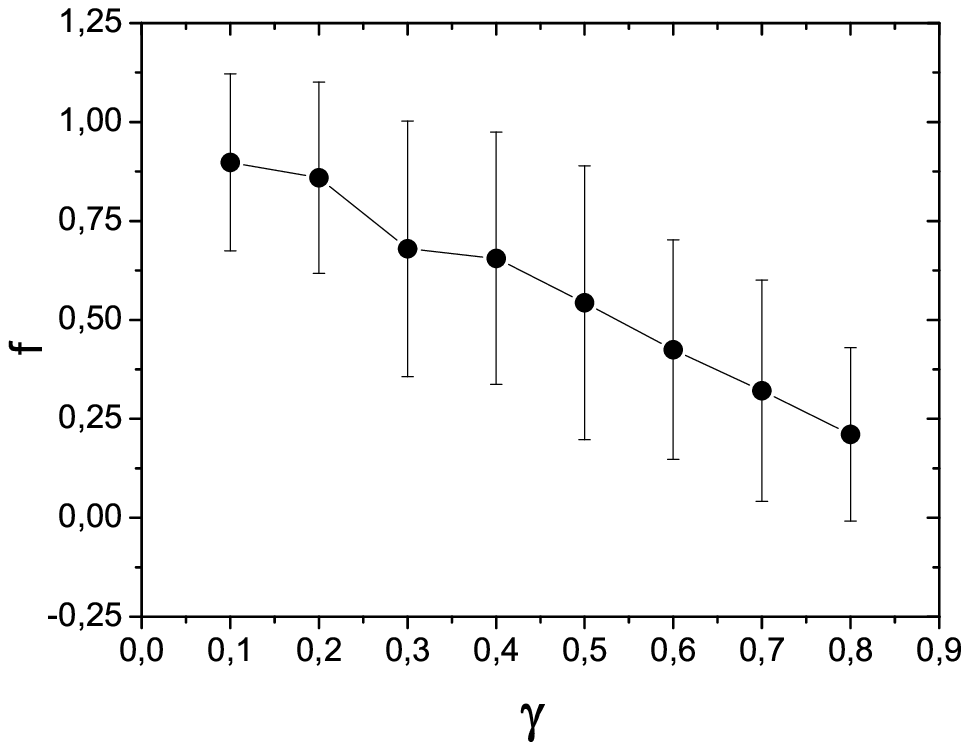}}
    \subfigure[]{\includegraphics[width=0.45\columnwidth]{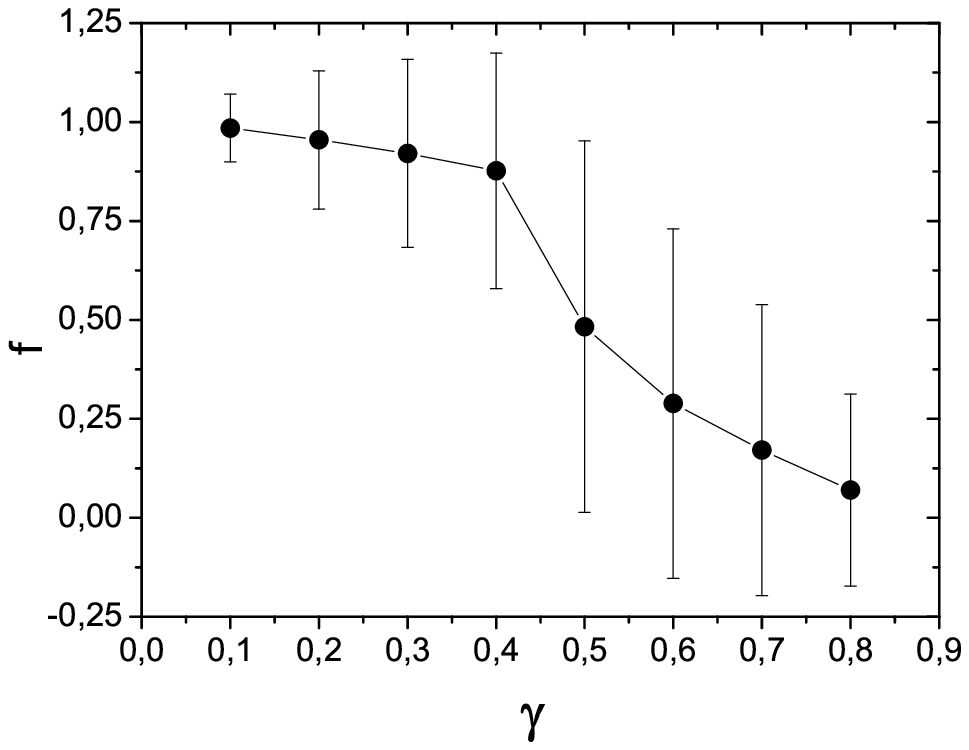}}
    \subfigure[]{\includegraphics[width=0.45\columnwidth]{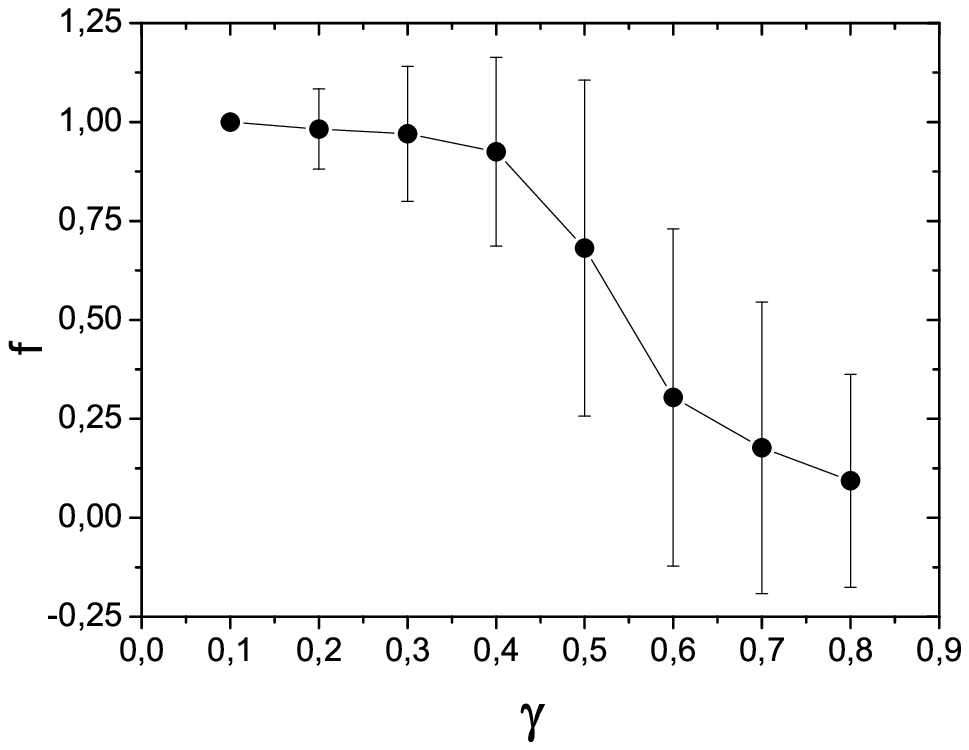}}
  \end{center}
  \caption{The average (and standard deviations) of the flag $f$
    indicating that the correct source has been identified among the
    detected evanescent trails with minimal overlap error $\xi$ for
    ER~(a), SW~(b), BA~(c) and DMS~(d) network models.}
  \label{fig:fevamodels}
\end{figure}

Figure~\ref{fig:fevamodels} shows the average and standard deviation of the values of the flag $f$ in terms
of $\gamma$ obtained for the same models.  Again, the results are similar to those obtained for the
Poissonian trails (Fig.~\ref{fig:fmodels}), but with a more gradual decrease of $f$ for the ER model.

Remarkably, though retaining less information about the original trail than the respectively Poissonian
counterparts, the evanescent trails tend to allow a similar identification of the source of the trail and the
original trail.

\subsection{Real networks}

\begin{figure}
  \begin{center}
    \subfigure[]{\includegraphics[width=0.45\columnwidth]{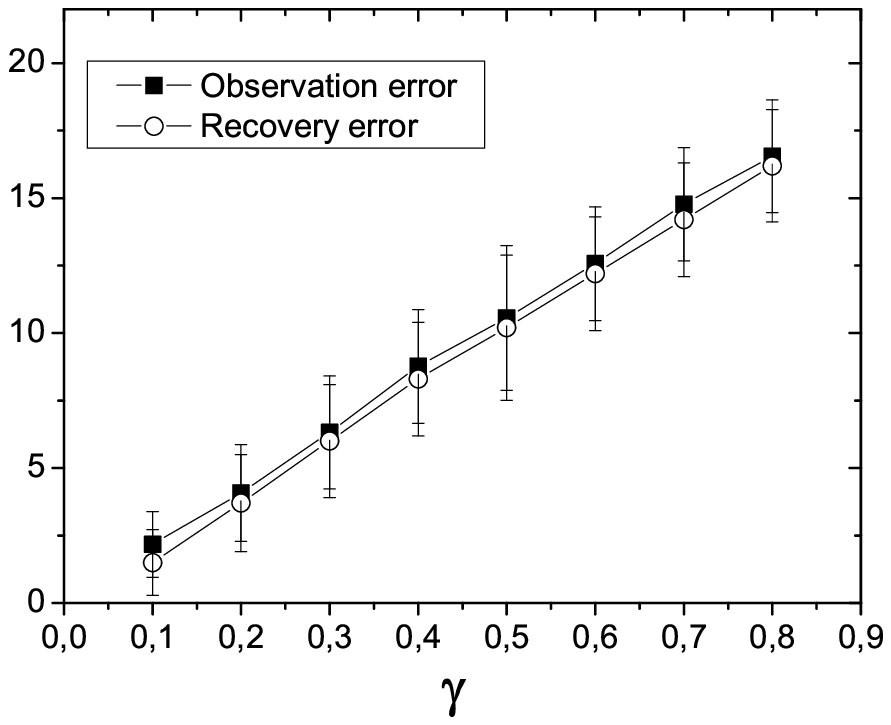}}
    \subfigure[]{\includegraphics[width=0.45\columnwidth]{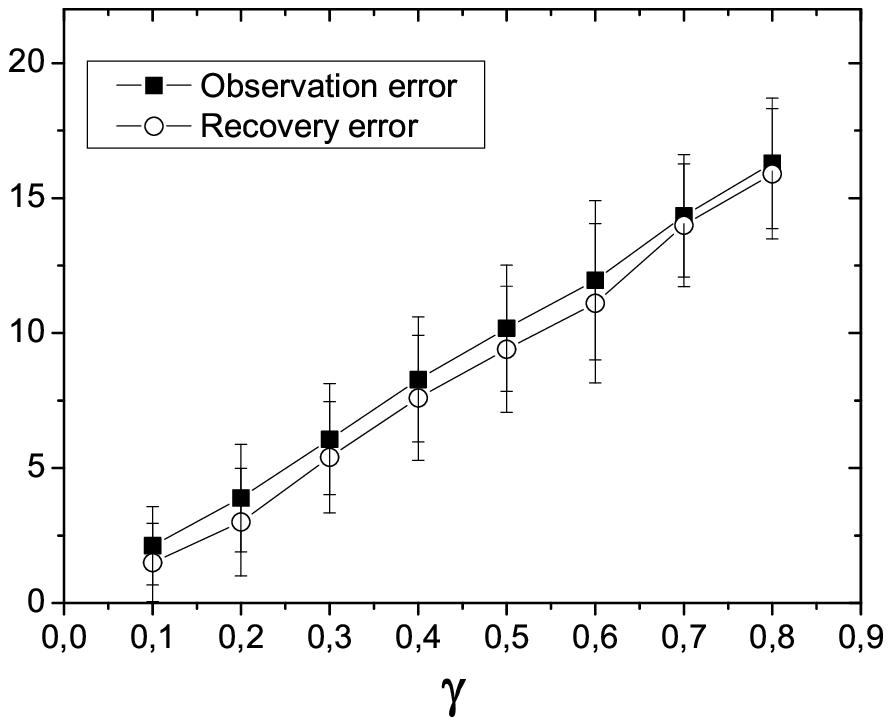}}
    \subfigure[]{\includegraphics[width=0.45\columnwidth]{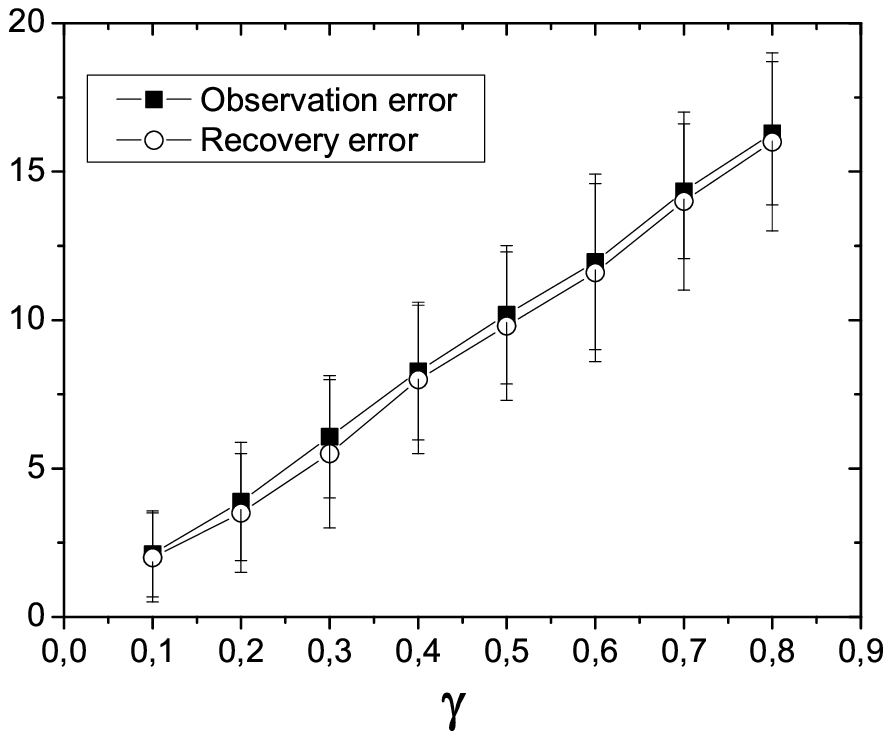}}
    \subfigure[]{\includegraphics[width=0.45\columnwidth]{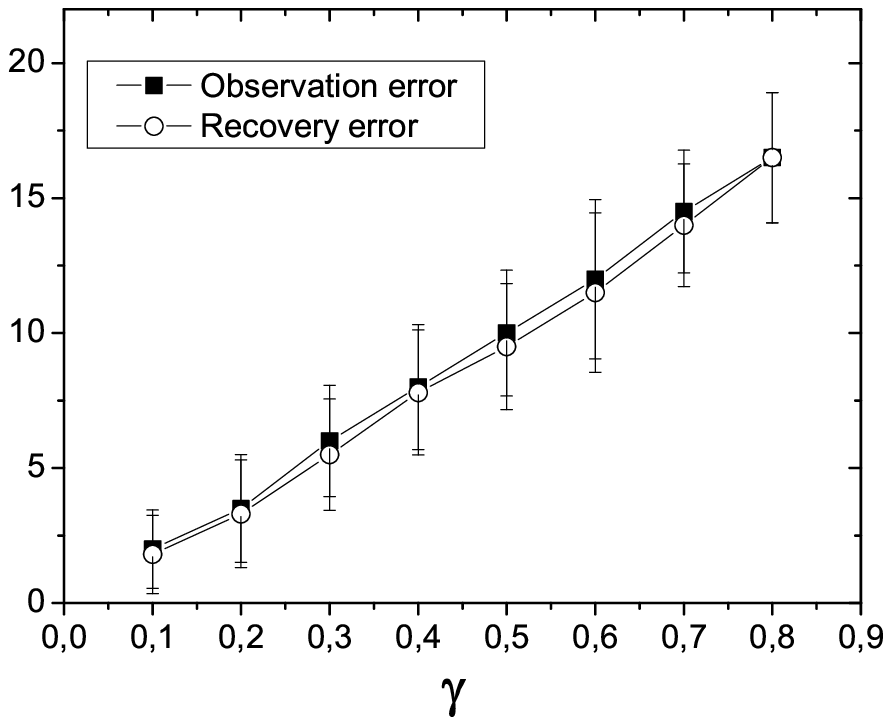}}
  \end{center}
  \caption{The observation (black squares) and recovery (white
    circles) errors obtained by using the recovery algorithm for
    Poissonian trail from random walks, for (a)~the Internet, (b)~the
    USA airlines, (c)~the e-mail network from the Univeristy Rovira i
    Virgili, and (d)~the scientific collaboration of complex networks
    researchers.}
  \label{fig:rec_nets}
\end{figure}

We considered four different networks in our simulations, namely: the Internet at the level of autonomous
systems, the US Airlines~\cite{pajek-data}, the e-mail network from the University Rovira i Virgili
(Tarragona)~\cite{guimera2003ssc} and the scientific collaboration of complex networks researchers.
Table~\ref{table1} presents some information about these networks. All random walk trails were Poissonian
with real extent equal to 20 nodes and all dilation trails took place along 2 hierarchies, with $\gamma=0.1,
0.2, \ldots, 0.8$. Figure~\ref{fig:rec_nets} shows average recovery errors, obtained for the Poissonian
random walk trails in the four considered real networks. Again, as we observed for the networks models, the
recovery error increases almost linearly with $\gamma$, being only slightly smaller than the observation
error.  The adopted recovery method achieves slightly better results for the US airlines network than for the
other networks.

\begin{table}
  \caption{Statistical measurements for the considered real
    networks. $N$ is the number of nodes, $\langle k \rangle$ is the
    average degree, $cc$ is the average clustering coefficient.}
\begin{tabular}{cccc}
  \hline
\textbf{Network}  & $N$ &$\langle k \rangle$ &$cc$ \\
  \hline
    Internet                                    &3,522   &3.59     &0.19 \\
    USA Airlines network                        &332     &12.81    &0.62 \\
    Collaboration in science                    &1,589   &3.45     &0.02 \\
    E-mail network                              &1,133   &19.24    &0.19 \\ \hline
\end{tabular}\label{table1}
\end{table}

\begin{figure}[t]
  \begin{center}
    \subfigure[]{\includegraphics[width=0.45\columnwidth]{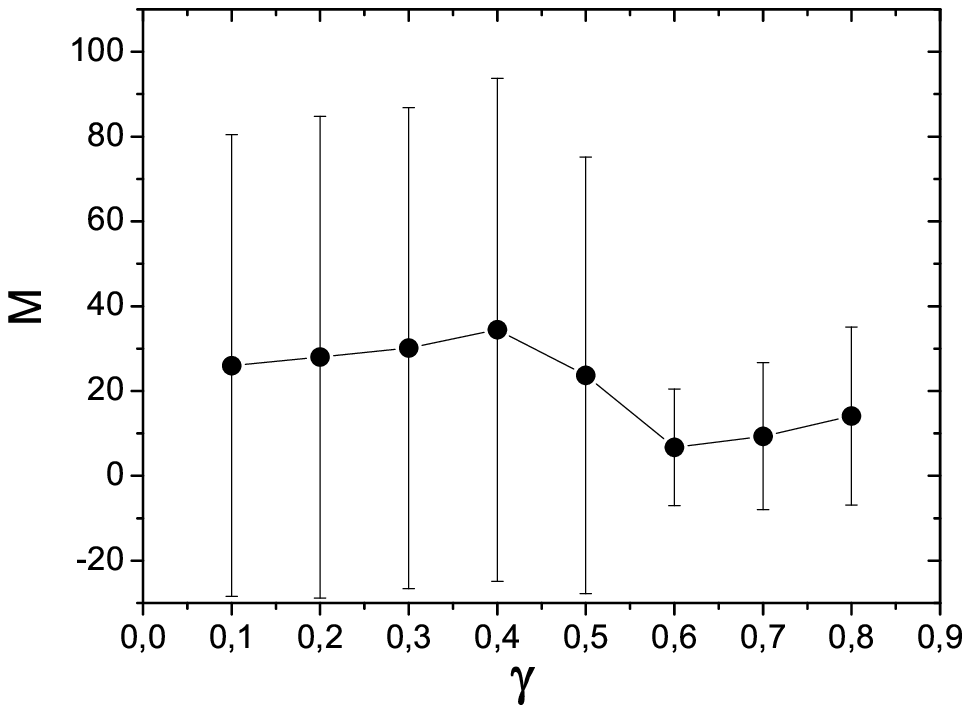}}
    \subfigure[]{\includegraphics[width=0.45\columnwidth]{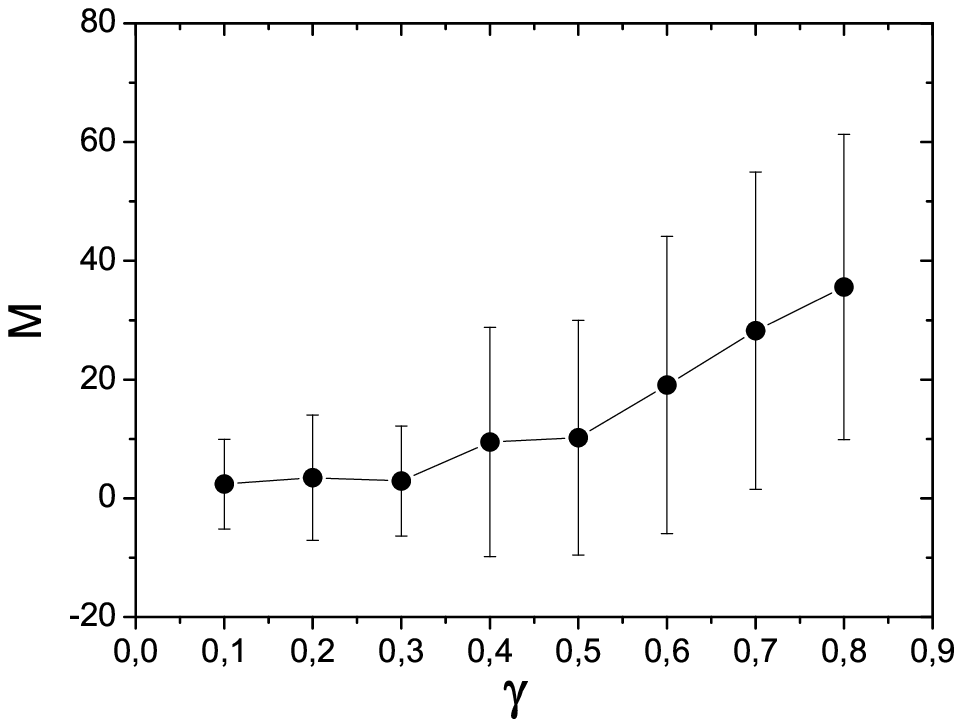}}
    \subfigure[]{\includegraphics[width=0.45\columnwidth]{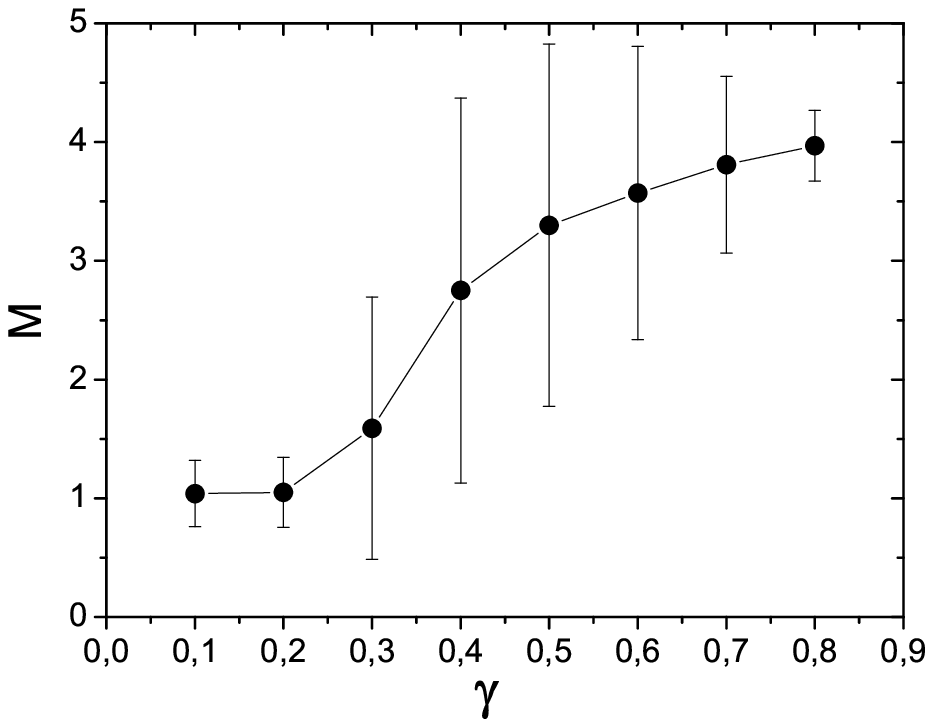}}
    \subfigure[]{\includegraphics[width=0.45\columnwidth]{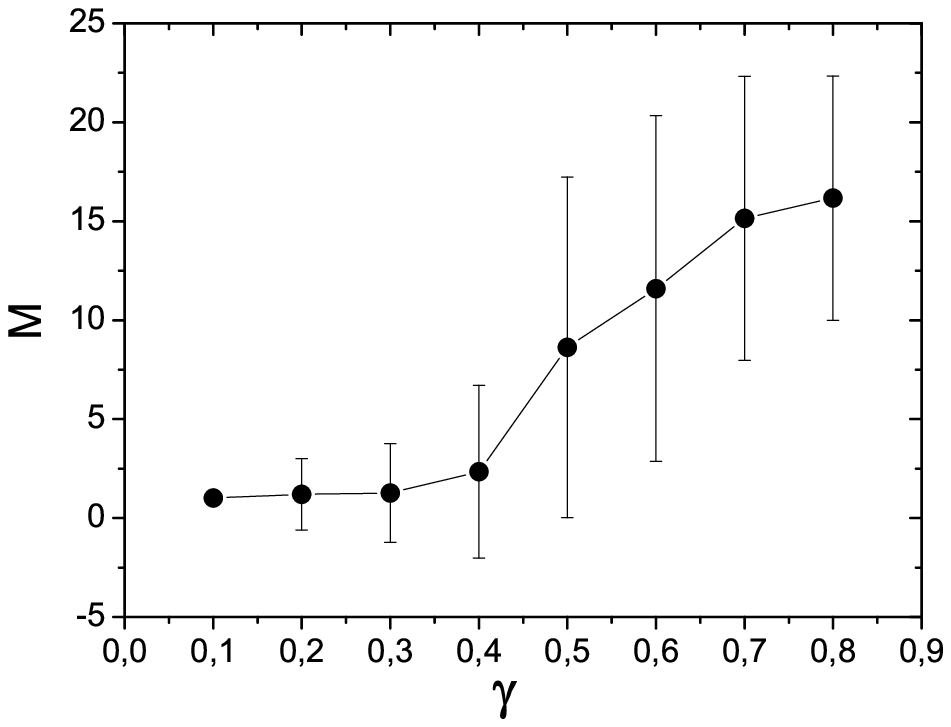}}
  \end{center}
  \caption{The average and standard deviations, in terms of $\gamma$,
    of the number $M$ of detected trails corresponding to the minimal
    overlap error obtained in the case of Poissonian dilation trails
    for (a)~the Internet, (b)~the USA airlines, (c)~the e-mail network
    from the Univeristy Rovira i Virgili, and (d)~the scientific
    collaboration of complex networks researchers.}
  \label{fig:M_nets}
\end{figure}

Figure~\ref{fig:M_nets} gives the average and standard deviation of $M$ for trails corresponding to the
minimal overlap error $\xi$ for Poissonian dilation trials in the considered real networks. The value of
$\left<M \right>$ tends to increase with $\gamma$ for all networks. For the Internet, $\left<M \right>$ has
two distinct behavior: (i)~for $\gamma \lesssim 0.4$ and $\gamma \gtrsim 0.6$ , $\left<M \right>$ increases
slowly, (ii)~for $0.4 \lesssim \gamma \lesssim 0.6$, $\left<M \right>$ decreases; in the region $\gamma
\lesssim 0.5$, $M$ has high standard deviations.  In the case of the US Airlines and the scientific
collaboration network, $\left<M \right>$ has a similar behavior, but has larger values than from the US
Airlines. The smallest values of $\left<M \right>$ are obtained for the e-mail network. Therefore, trails can
be better recovery in this type of network, which is an important discovery because it has implications for
the identification of the source of spreading of virus or rumors, among other cases.

\begin{figure}
  \begin{center}
    \subfigure[]{\includegraphics[width=0.45\columnwidth]{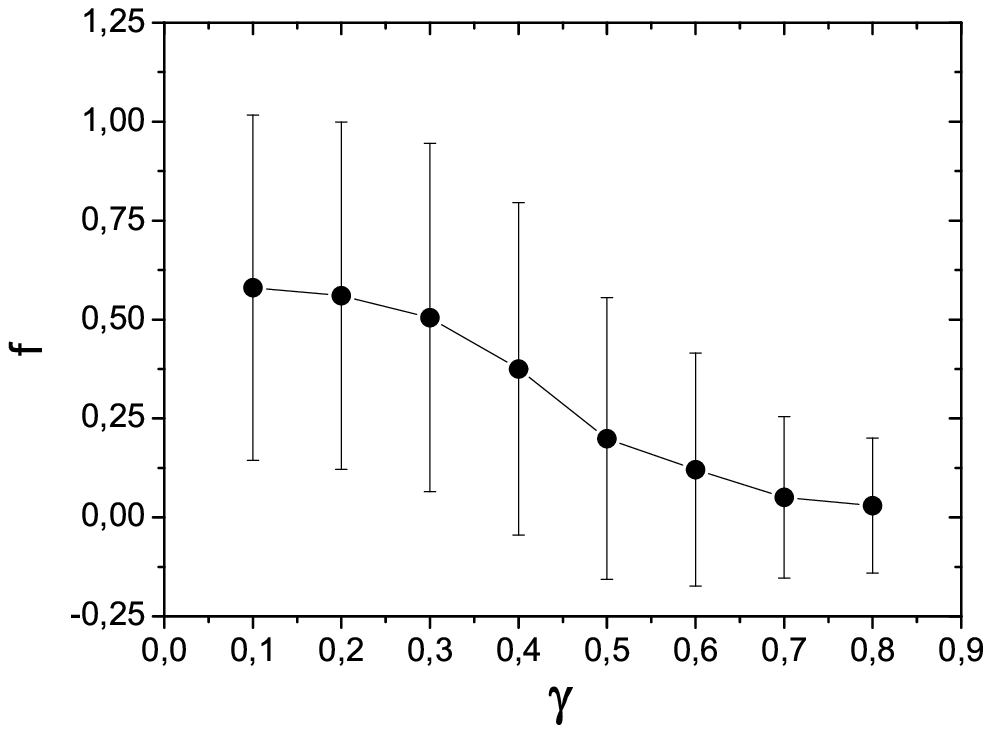}}
    \subfigure[]{\includegraphics[width=0.45\columnwidth]{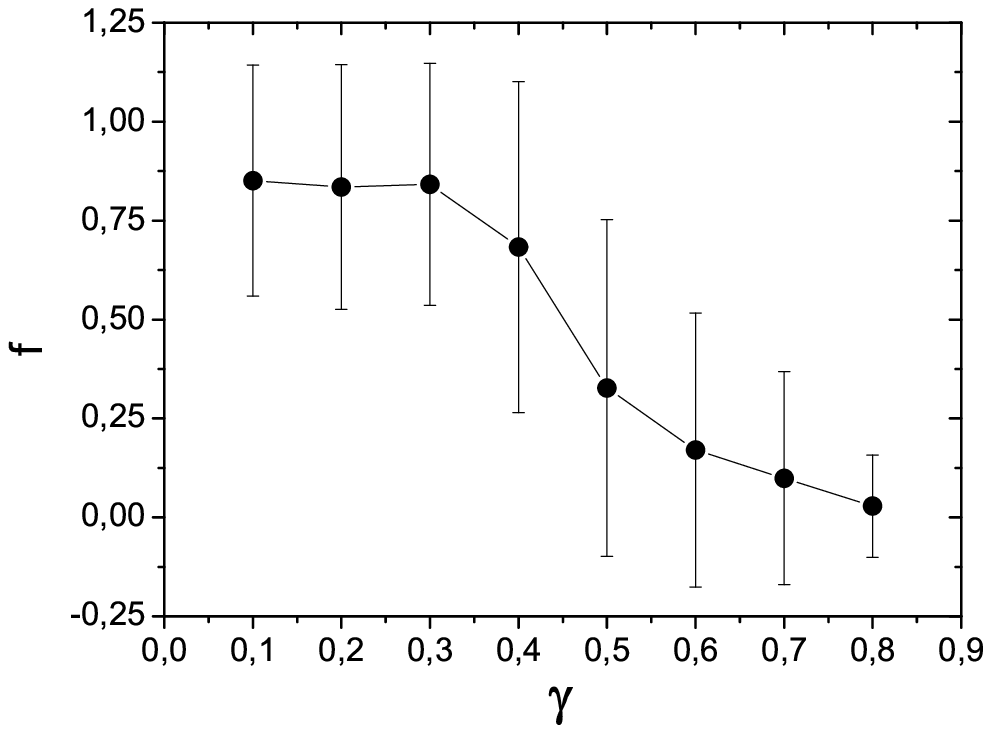}}
    \subfigure[]{\includegraphics[width=0.45\columnwidth]{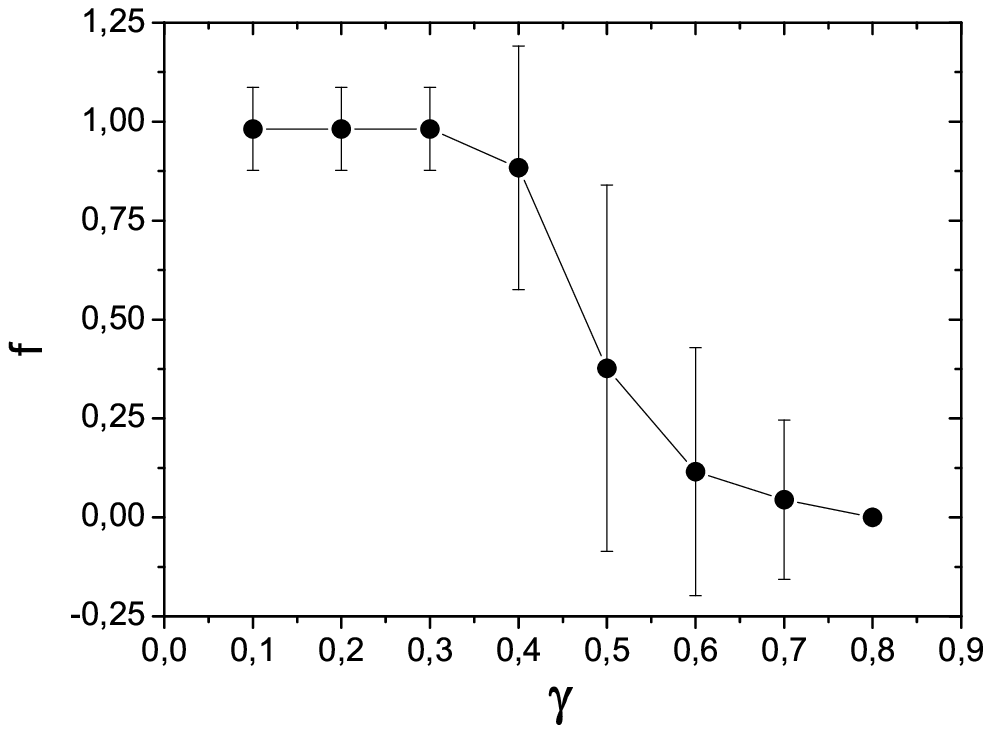}}
    \subfigure[]{\includegraphics[width=0.45\columnwidth]{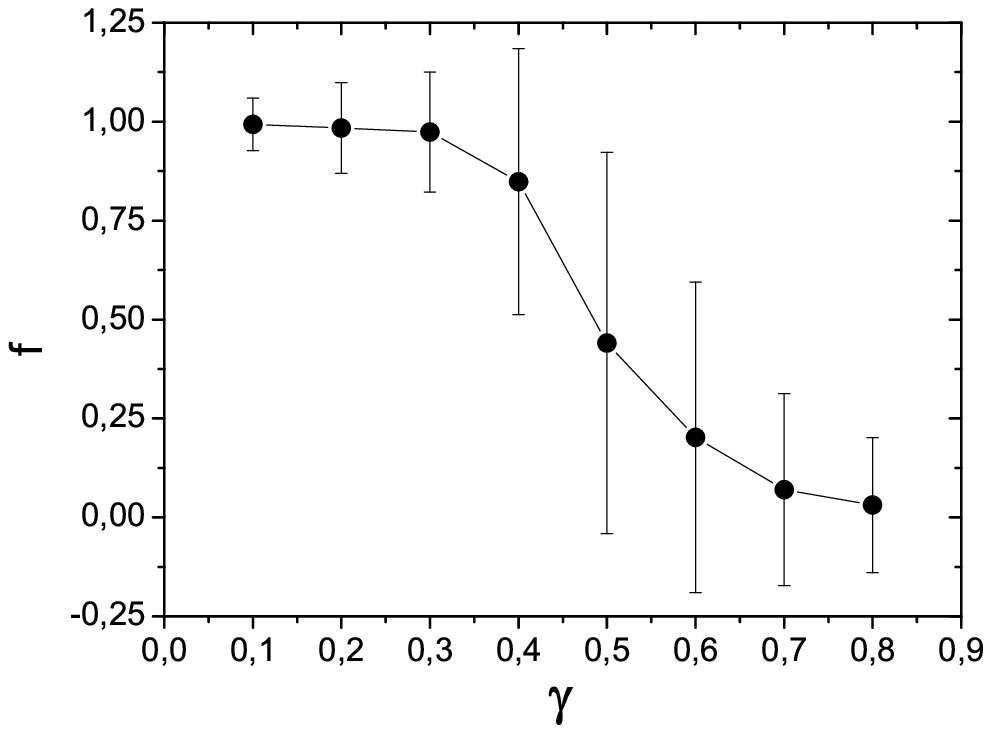}}
  \end{center}
  \caption{The average (and standard deviations) of the flag $f$
    indicating that the correct source has been identified among the
    detected Poissonian trails with minimal overlap error $\xi$
    (a)~the Internet, (b)~the USA airlines, (c)~the e-mail network
    from the Univeristy Rovira i Virgili, and (d)~the scientific
    collaboration of complex networks researchers.}
  \label{fig:f_nets}
\end{figure}

The average $\left< f \right>$ of the correct source identification flag (and standard deviation) is given in
terms of $\gamma$ in Figure~\ref{fig:f_nets} for the considered real networks. The source identification is
worst for the Internet.

\begin{figure}
  \begin{center}
    \subfigure[]{\includegraphics[width=0.45\columnwidth]{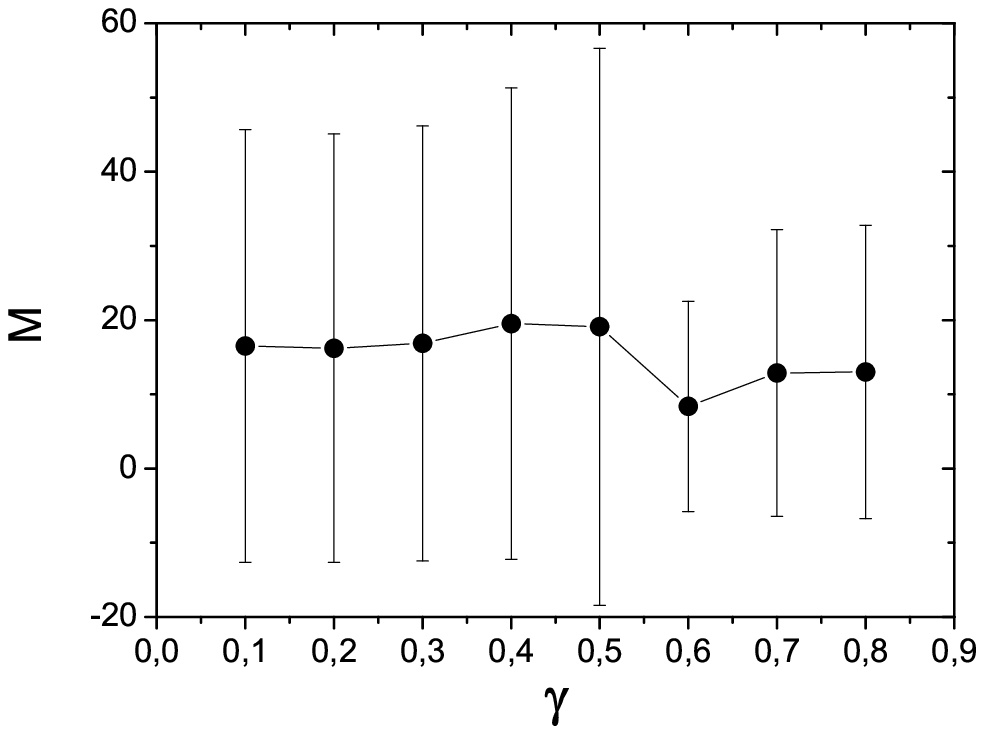}}
    \subfigure[]{\includegraphics[width=0.45\columnwidth]{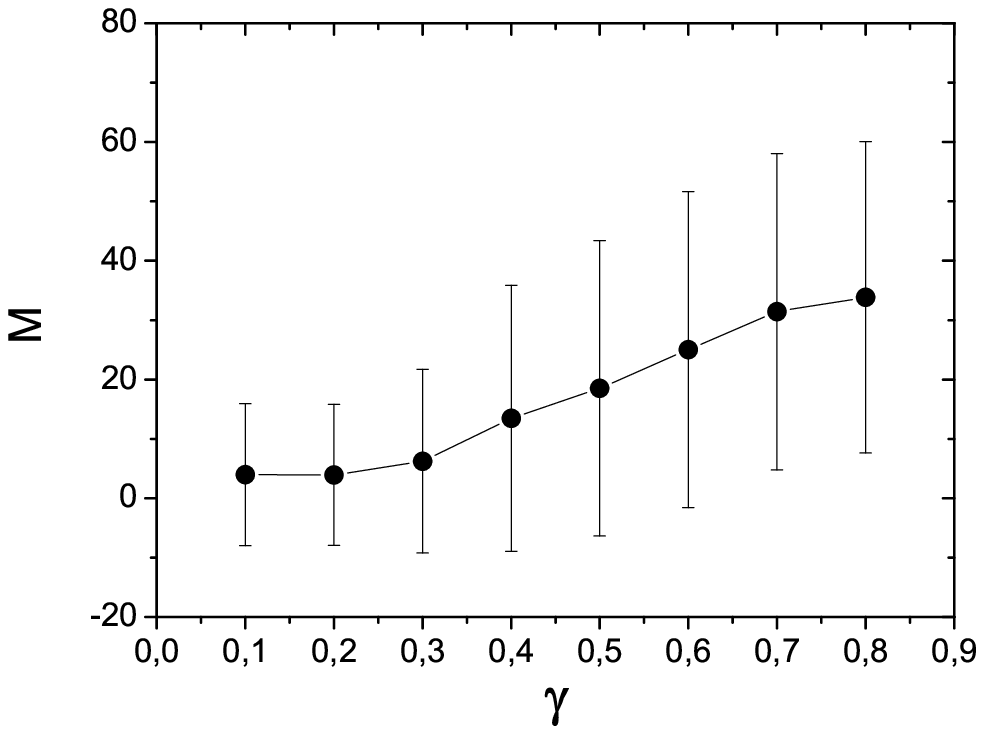}}
    \subfigure[]{\includegraphics[width=0.45\columnwidth]{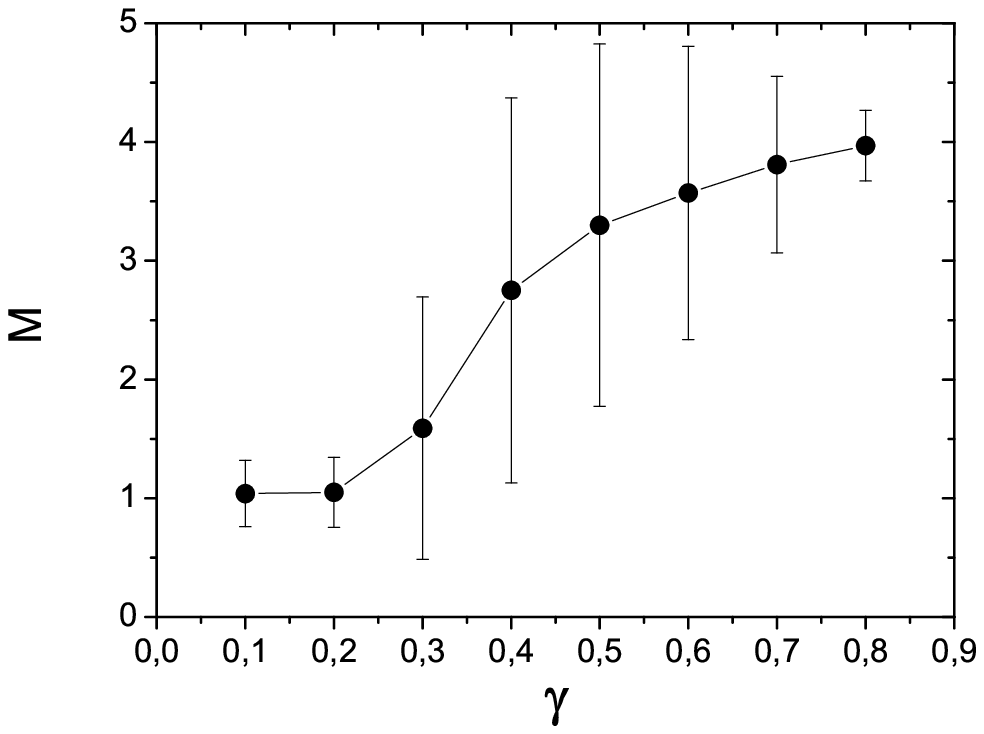}}
    \subfigure[]{\includegraphics[width=0.45\columnwidth]{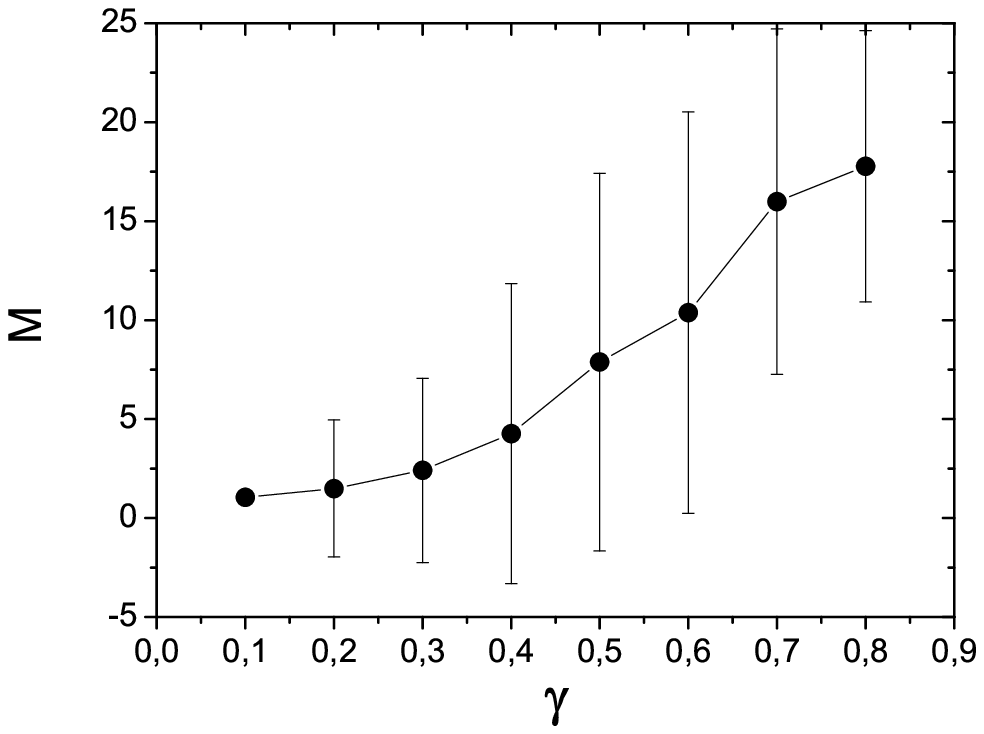}}
  \end{center}
  \caption{The average and standard deviations, in terms of $\gamma$,
    of the number $M$ of detected evanescent trails corresponding to
    the minimal overlap error obtained (a)~the Internet, (b)~the USA
    airlines, (c)~the e-mail network from the Univeristy Rovira i
    Virgili, and (d)~the scientific collaboration of complex networks
    researchers.}
  \label{fig:Meva_nets}
\end{figure}

\begin{figure}[t]
  \begin{center}
    \subfigure[]{\includegraphics[width=0.45\columnwidth]{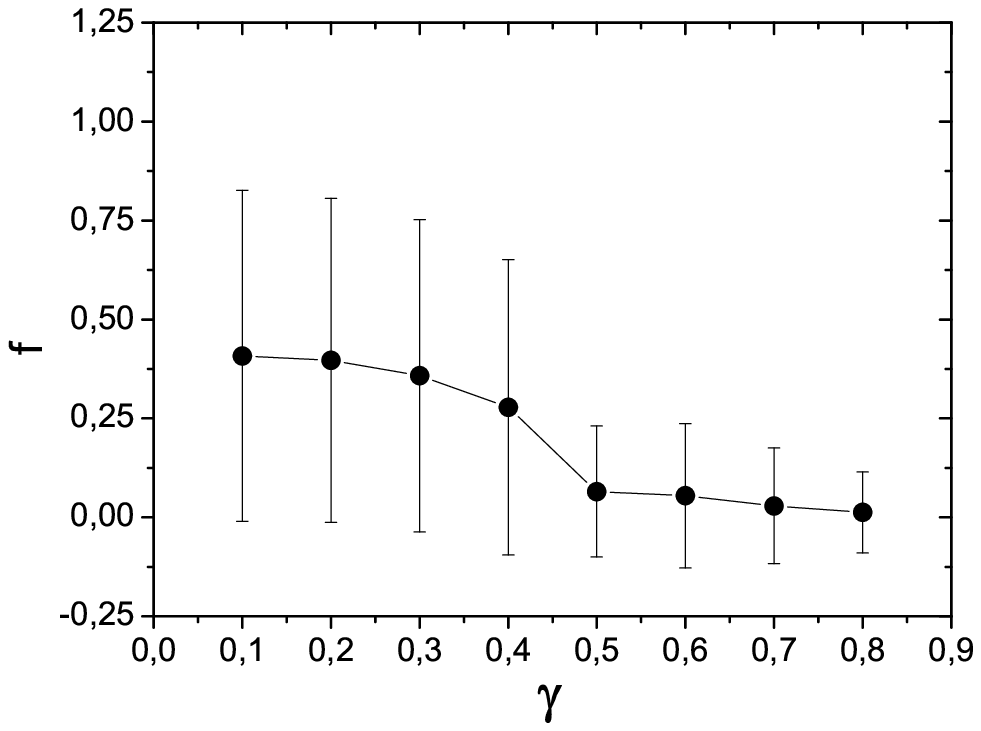}}
    \subfigure[]{\includegraphics[width=0.45\columnwidth]{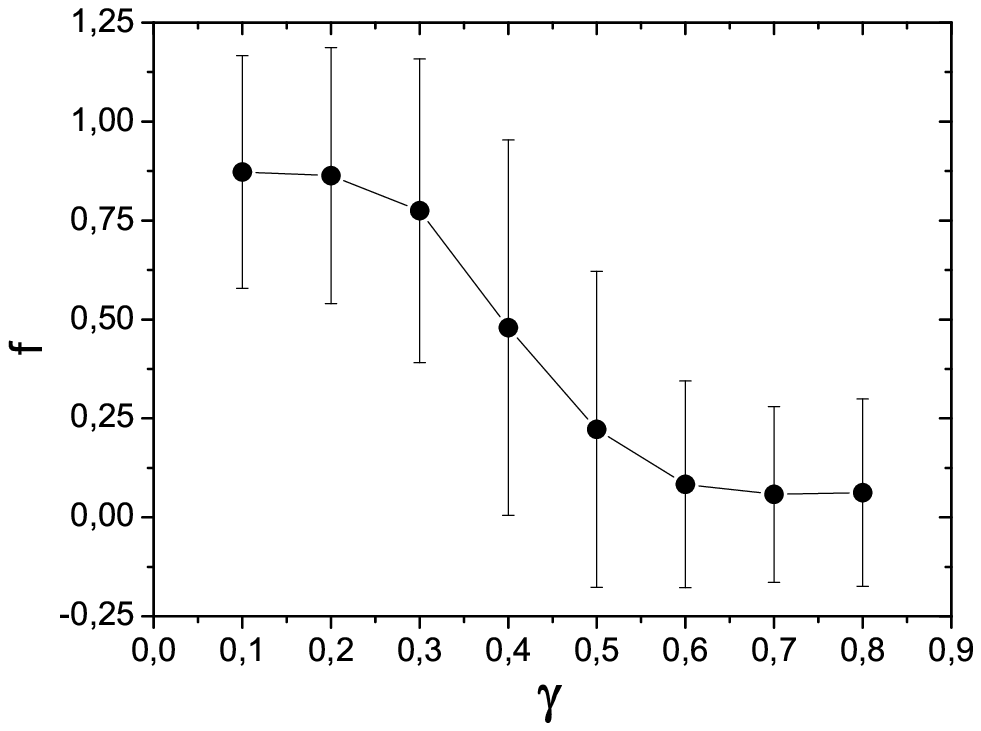}}
    \subfigure[]{\includegraphics[width=0.45\columnwidth]{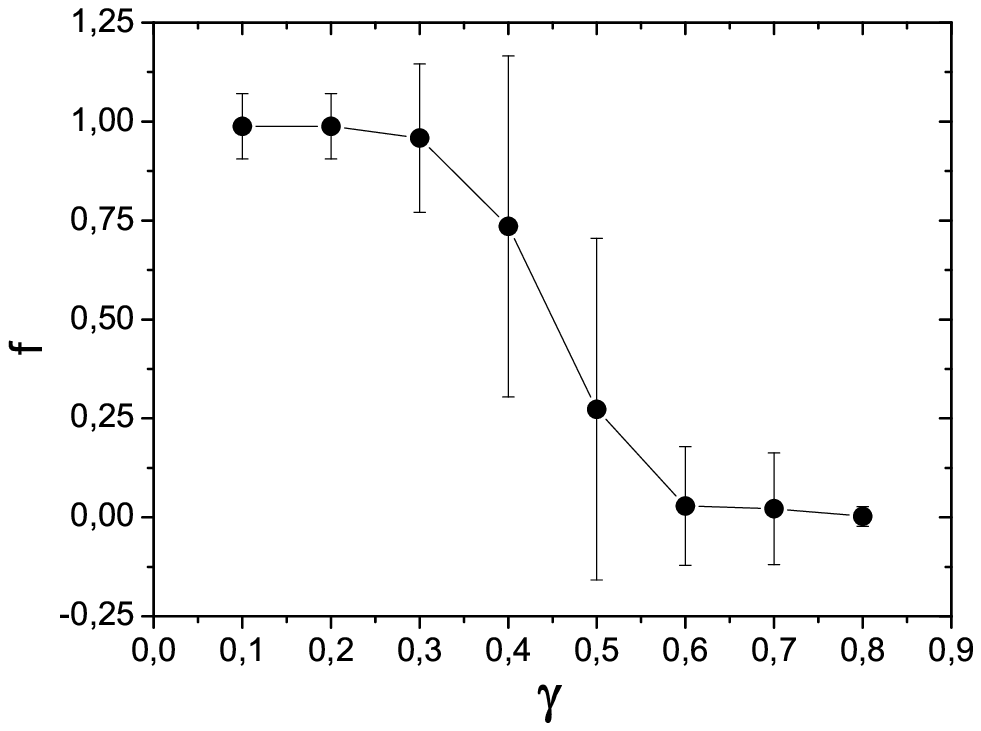}}
    \subfigure[]{\includegraphics[width=0.45\columnwidth]{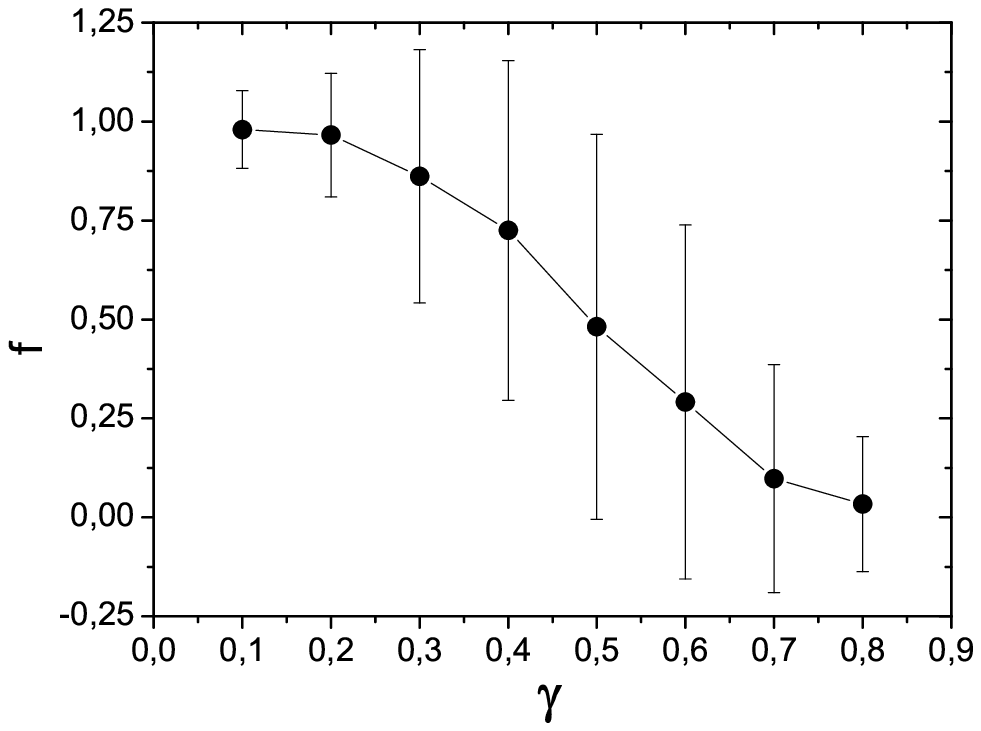}}
  \end{center}
  \caption{The average (and standard deviations) of the flag $f$
    indicating that the correct source has been identified among the
    detected evanescent trails with minimal overlap error $\xi$
    (a)~the Internet, (b)~the USA airlines, (c)~the e-mail network
    from the Univeristy Rovira i Virgili, and (d)~the scientific
    collaboration of complex networks researchers.}
  \label{fig:feva_nets}
\end{figure}

For transient dilation trails of the \emph{evanescent} category, the results are shown in
Figure~\ref{fig:Meva_nets} (for $M$) and Figure~\ref{fig:feva_nets} (for $f$).  As for the models, the
results are close to those obtained considering Poissonian dilation trails, despite the fact that the
evanescent category provides less information for trail recovery.

\section{Multi-agents}\label{Sec:agents}

\begin{figure}
  \begin{center}
    \subfigure[]{\includegraphics[width=0.45\columnwidth]{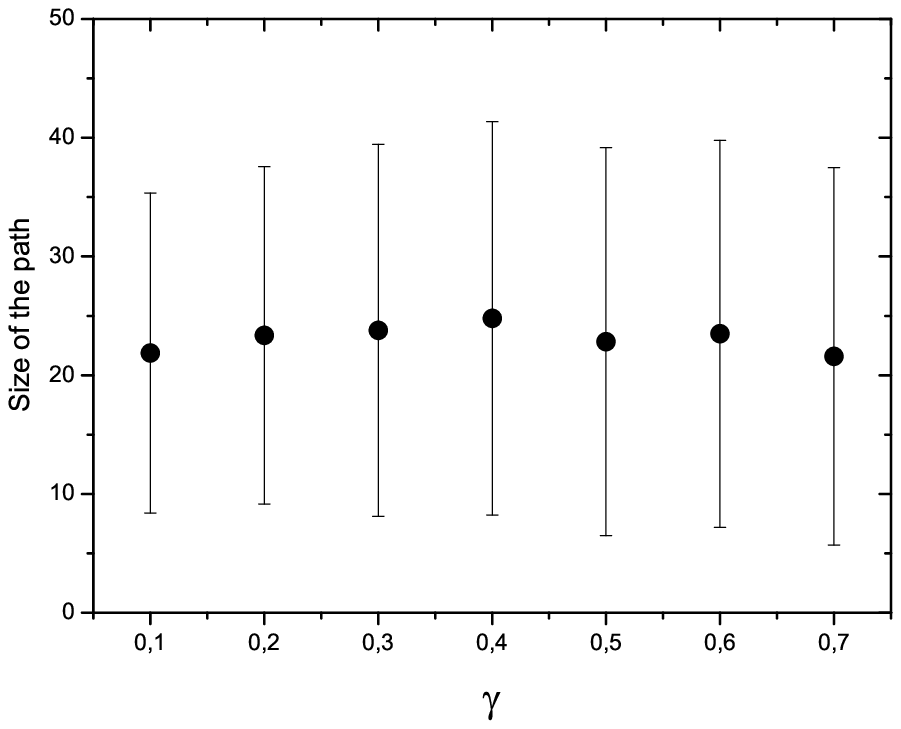}}
    \subfigure[]{\includegraphics[width=0.45\columnwidth]{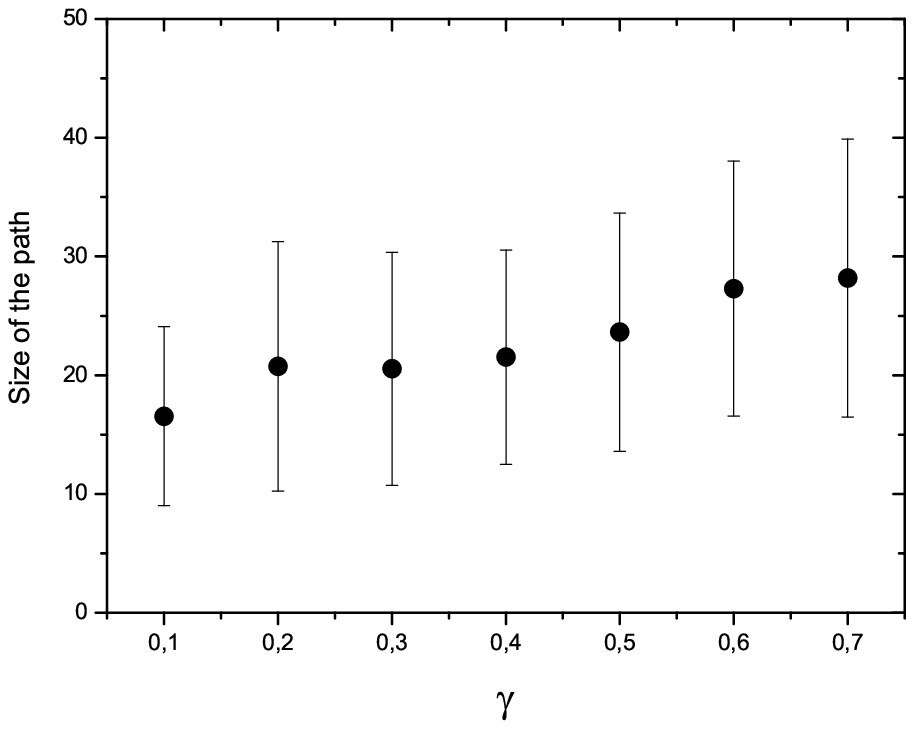}}
    \subfigure[]{\includegraphics[width=0.45\columnwidth]{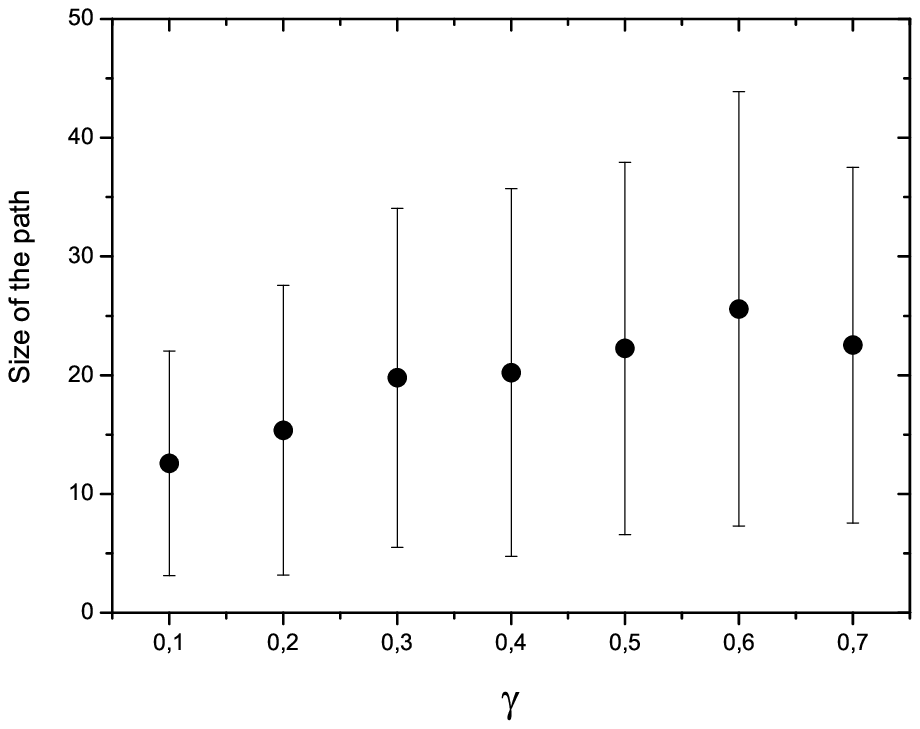}}
    \subfigure[]{\includegraphics[width=0.45\columnwidth]{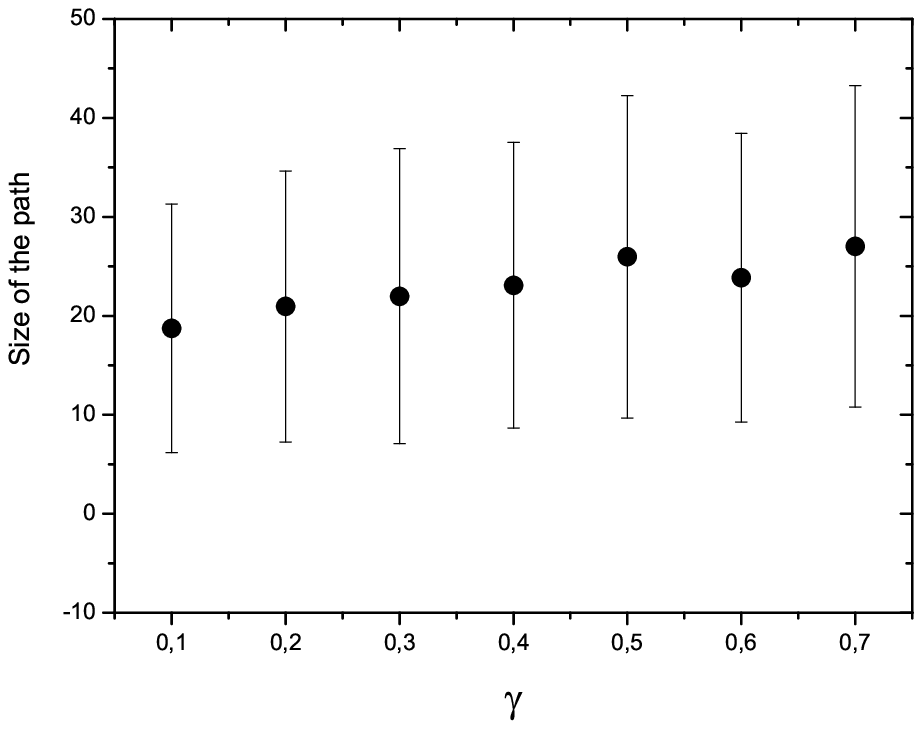}}
  \end{center}
  \caption{The average and standard deviations in terms of the length of
  the path covered by the second agent obtained for ER~(a), SW~(b),
  BA~(c) and DMS~(d) network models.  Each point is an average of 500
  realizations.}  \label{fig:Agents}
\end{figure}

We considered the dynamics of multi-agents on trail evolution
considering four complex networks models: ER, SW, BA and DMS. Each
considered network model is formed by $N = 1\,000$ nodes and average
degree $\left< k \right> = 4$. The process is defined as follows: (i)
the first agent leaves a gradient trail --- the current position has
the strongest mark and the source, the weakest --- by self avoid
random walks, (ii) the path is erased with a probability $\gamma$
(Poissonian trail as before), (iii) the second agent tries to reach
the target (the last vertex of the trail) by following preferentially
the strongest, at each immediate neighborhood, the marks left by the
first agent. When the second agent does not find any mark, it performs
a random walk until another mark is found. This process is performed
for example by ants in searching of food --- the first agent can
represent an ant that leaves a trail of pheromone that will be
followed by the second ant. The objective of our investigation is to
determine the influence of the topology in target identification
efficiency, as well as possible overall trajectory minimization, by
measuring the length of the path covered by the second agent. All random
walk trails were Poissonian with real extent equal to 20 nodes and
$\gamma=0.1, 0.2, \ldots, 0.8$. Figure~\ref{fig:Agents} presents the
length of the path covered by the second agent in function of the
erasing rate $\gamma$. As can be clearly seen, when $\gamma < 0.5$ the
second agent covers smallest paths for BA, SW and DMS network models,
followed by the ER. This suggests that the topology of the network is
fundamental for trajectory following. Indeed, the hubs present in BA
and DMS network models provide shortcuts through the network. Enhanced
efficiency was also found for the SW network models, but the high
clustering coefficient was identified as being fundamental in this
case.  While the length of the path obtained by the second agent is
kept almost constant when $\gamma$ increases for ER, it increases in
the remainder models. For $\gamma > 0.5$, the length of the path for
ER reaches its smallest value. Therefore, when the trail is almost
complete, the BA, SW and DMS topologies provide the best performances,
but when the trail is sparse, ER allows the shortest paths. Thus, the
topology was verified to strongly influence agent dynamics.

\section{Concluding Remarks}

Great part of the interest in complex networks has stemmed from their
ability to represent and model intricate natural and human-made
structures ranging from the Internet to protein interaction
networks. There is a growing interest in the study of dynamics in such
systems (e.g., \cite{Newman03:SIAM, Boccaletti:2006,Costa_know:2006}).
Among the many types of interesting dynamics which can take place on
complex networks, we have the evolution of trails left by moving
agents during random walks and dilations. In particular, given one of
such (possibly incomplete) trails, immediately implied problems
involve the recovery of the full trail and the identification of its
possible source.  Such problems are particularly important because
they are directly related to a large number of practical and
theoretical situations, including fad and rumor spreading,
epidemiology, exploration of new territories, transmission of messages
in communications, amongst many other possibilities.

The important problem of analyzing trails left in networks by moving
agents during random walks and dilations has been formalized and
investigated by using two heuristic algorithms in the present
article. We considered four models of complex networks, namely
Erd\H{o}s-R\'enyi, Barab\'asi-Albert, Watts-Strogatz, and
Dorogovtsev-Mendes-Samukhin models, and four different real networks:
the Internet at the level of autonomous systems, the US Airlines, the
e-mail network from the University Rovira i Virgili (Tarragona) and
the scientific collaboration of complex networks researchers.  Also,
we considered two types of trails: permanent and transient. Particular
attention was given to trails with transient marks.  In the case of
random walk trails, we investigated how incomplete Poissonian trails
can be recovered by using a shortest path approach.  The recovery and
identification of source of dilation trails was approached by
reproducing the dilating process for each of the network nodes and
comparing the obtained trails with the observable state variables.

It has been shown through simulation that both such strategies are
potentially useful for trail reconstruction and source
identification. In addition, a series of interesting results and
trends have been identified.  First, it has been found that the
shortest path approach for recovery of trails left by random walks
provides similar results for all considered networks and network
models, which suggests that such strategy independes on the network
topology. Second, for dilatation trails it was found that the
Poissonian and evanescent types of trails allow similar efficiency in
the identification of sources, despite the fact that the latter trails
incorporate less information than the former.

The analysis of multi-agents on networks showed that the topology
strongly influences the respective performance. When the trail is
almost complete, the Barab\'asi-Albert, Watts-Strogatz and
Dorogovtsev-Mendes-Samukhin network models provide the best
performance. On the other hand, when the information about the trail
is sparse, the final point of the trail is reached faster for the
Erd\H{o}s-R\'enyi network model.

It is believed that the suggested methods and experimental results
have paved the way to a number of important related works, including
the investigation of the scaling of the effects and trends identified
in the present work to other network sizes, average node degrees and
network models.  At the same time, it would be interesting to consider
graded state variables, more than a single trail taking place
simultaneously in a network, other types of random walks (e.g.,
preferential~\cite{Costa_know:2006}), as well as alternative recovery
and source identification strategies. One particularly promising
future possibility regards the recovery of diffusive dynamics in
complex networks.

\section{Acknowledgments}
Luciano da F. Costa is grateful to CNPq (308231/03-1 and 301303/06-1)
and FAPESP (05/00587-5) for financial support.  Francisco A. Rodrigues
acknowledges FAPESP sponsorship (proc. 04/00492-1).

\bibliographystyle{apsrev}
\bibliography{trails}
\end{document}